\documentclass{emulateapj}
\usepackage{amssymb}
\usepackage{apjfonts}
\usepackage{epsfig}
\usepackage{natbib}
\bibpunct{(}{)}{,}{a}{}{}

\newcommand*{\Msun}{\ensuremath{\mathrm{M_\odot}}}%
\newcommand*{\ML}{\ensuremath{M/L}}%
\newcommand{\sersic}{S\'ersic}

\def\deg      {{\ifmmode^\circ\else$^\circ$\fi}} 

\def\lsim{\mathrel{\rlap{\lower4pt\hbox{\hskip1pt$\sim$}}\raise1pt\hbox{$<$}}}
\def\gsim{\mathrel{\rlap{\lower4pt\hbox{\hskip1pt$\sim$}}\raise1pt\hbox{$>$}}}   

\slugcomment{}

\shorttitle{Early and late type galaxies in the COSMOS}
\shortauthors{Pannella et al.}

\begin{document}


\title{The evolution of early and late type galaxies in the COSMOS up to $z\approx1.2$\altaffilmark{*}}


\author{Maurilio Pannella\altaffilmark{1,2}, Armin Gabasch\altaffilmark{1,3}, Yuliana Goranova\altaffilmark{4,5,6}, Niv Drory\altaffilmark{1}, Ulrich Hopp\altaffilmark{1,4}, Stefan Noll\altaffilmark{1,7}, Roberto~P.~Saglia\altaffilmark{1}, Veronica Strazzullo\altaffilmark{2} and Ralf Bender\altaffilmark{1,4}}


\altaffiltext{*}{Based on observations made with the Advanced Camera for Surveys on board the NASA/ESA {\it Hubble Space Telescope} (GO Proposal 9822)}
\altaffiltext{1}{Max-Planck-Institut f\"ur extraterrestrische Physik, Giessenbachstr., Postfach 1312, D-85741 Garching bei  M\"unchen, Germany.}
\altaffiltext{2}{National Radio Astronomy Observatory, P.O. Box 0, Socorro, NM 87801-0387;~mpannell@nrao.edu~.}
\altaffiltext{3}{European Southern Observatory, Karl Schwarzschild Strasse 2,  Garching bei  M\"unchen, Germany.}
\altaffiltext{4}{Universit\"atssternwarte M\"unchen, Scheinerstr. 1, D-81673 M\"unchen, Germany.}
\altaffiltext{5}{Institut d'Astrophysique de Paris, 98 bis boulevard Arago, F-75014 Paris, France.}
\altaffiltext{6}{Institute of Astronomy, Bulgarian Academy of Sciences, 72 Tsarigradsko Chaussee Blvd., 1784 Sofia, Bulgaria.}
\altaffiltext{7}{Observatoire Astronomique de Marseille-Provence (OAMP), 38 rue Fr\'ed\'eric Joliot-Curie, 13388 Marseille cedex 13, France.}


\begin{abstract}
The Cosmic Evolution Survey (COSMOS) allows for the first time a
highly significant census of environments and structures up to
redshift one, as well as a full morphological description of the
galaxy population.  In this paper we present a study aimed to
constrain the evolution, in the redshift range $0.2 < z < 1.2$, of the mass content of
different morphological types and its dependence on the environmental
density.  We use a deep multicolor catalog, covering an area of
$\sim$0.7 $\sq \degr$ inside the COSMOS field, with accurate
photometric redshifts (i $\lsim$ 26.5 and
\mbox{$\Delta z / (z_{spec}+1) \approx 0.035 $}). We estimate galaxy
stellar masses by fitting the multi-color photometry to a grid of
composite stellar population models. We quantitatively describe
the galaxy morphology by fitting PSF convolved \sersic~profiles to the
galaxy surface brightness distributions down to F814 = 24 mag for a sample 
of 41300 objects.

We confirm an evolution of the morphological mix with redshift: the
higher the redshift the more disk-dominated galaxies become important. We find that the
morphological mix is a function of the local comoving density: the morphology density relation extends up to the highest redshift explored.

The stellar mass function of disk-dominated galaxies is
consistent with being constant with redshift. Conversely, the stellar mass function of bulge-dominated systems shows a decline in normalization with
redshift. Such different behaviors of late-types and early-types stellar mass
functions naturally set the redshift evolution of the {\it transition} mass. 

We find a population of relatively massive, early--type galaxies, having
high SSFR and blue colors which live preferentially in low density environments. 
The bulk of massive ($> 7 \times 10^{10} M_\odot $)~early--type galaxies have similar characteristic ages, colors, and SSFRs independently of the environment they belong to,
with those hosting the oldest stars in the Universe preferentially belonging to the highest density regions.

The whole catalog including morphological information and stellar mass estimates 
analysed in this work, is made publicly available.

\end{abstract}





\section{Introduction}

Galaxy formation and evolution has been a very actively debated topic
of observational cosmology in the last years. Only recently models and
observations are converging toward a unified and coherent picture.

On one side the observations of the Universe, with some controversies
on the way, have finally agreed that half of the present-day stars was
already in place at z $\approx$ 1 and that the same mass deficit
affects, within the measured accuracies, the high-mass end: basically, the 
galaxy stellar mass function evolves from the local determination to redshift one only by a
normalization factor of about two. This means that, both at high and at
low masses, there is roughly a factor of two difference between the comoving number
density at redshift one and the local value
\citep[e.g.][]{dickinson2003,rudnick2003,drory2004,bundy2005,P06, borch06,font06,pozz07,arn07,pg2008,cowie2008,niv2008,marchesini2008,ilbert091}.
Measurements of the star formation rate density over cosmic times
\citep[e.g.][]{giavalisco2004, gabasch:sfr, bouw2004, saw06,bouw07} are also in good
agreement with half of the present-day stars being already born at
redshift 1.
 
On the other side the models, linking the hierarchical growth of dark
matter structures to the observed galaxy properties by means of
simplified prescriptions for the formation of baryonic systems,
predict that galaxies~form in a bottom-up fashion by following the
cosmological destiny of dark matter halos. Massive galaxies, in these
models, assemble most (50\%) of their stellar mass via merging at
$z<1$ ~\citep{gdl06}, but most of these accretion events are red
mergers (i.e. gas-free mergers that imply no induced star-formation),
which are very difficult to detect from the
observational point of view~\citep[e.g.][]{hop08,cox08}. The latest realizations of these models
\citep[e.g.][]{bower06,kitzb07,  catt08,font09} have been able to fully reproduce the
galaxy stellar mass function up to high redshifts, thus reconciling
the theoretical bottom-up assembly of dark matter halos to the claimed
top-down assembly of galaxies ~\citep[e.g.][]{cim06}.

In {\it hierarchical} models star formation is supposed to take place
only in disk structures and in gas-rich mergers~(see also \citealt{dekel2009} for an up to date perspective). This latter event
would exhaust the residual gas and create a dynamically hot system: a
bulge galaxy. Also dynamical instabilities and minor mergers are
able to destroy the disk structures and finally create an elliptical
galaxy. Detailed predictions in the literature do not exist but the
general expectation is that the fraction of bulge--dominated massive
galaxies increases with time.

The study of the evolution of the morphological mix in the galaxy
stellar mass function is a decisive tool to put sensitive constraints
on the models, because it could give unique insights in both the
feedback and the merging processes in galaxy evolution \citep[see for
a detailed discussion][]{cole2000}.

Galaxies have been often classified, both at low and at high redshift, as "red sequence" and "blue cloud" objects based on their broad band colors  
~\citep{baldry2004,fontana:2,giall05,bundy06}. However this kind of classification when used as a proxy for the galaxy morphology~(see e.g. \citealt{mobasher09}), suffers from  some obvious drawbacks: a disk galaxy populated by an old stellar population
would be classified as an early--type object, and vice versa; a starburst galaxy, heavily obscured by dust, would be classified as a red, and {\it dead}, early--type object.

Thanks to the advent of the Advanced Camera for Surveys
(ACS;~\citealt{ford03}) on board the Hubble Space Telescope~(HST), it has
become possible to investigate the optical rest-frame emission of
galaxies up to redshift one and slightly beyond with sub-kiloparsec
resolution and on relatively wide fields for hundreds of thousands of galaxies.

Given the exponentially growing number of galaxies available to be morphologically classified it has become urgent for galaxy evolution studies to build up robust and automated tools for morphological and structural classification that could reliably substitute the visual classification~(see the discussion in \citealt{bamford09} on unavoidable drawbacks affecting the automated approach). The efforts on the path to the automated image classification have essentially split in two main branches, parametric and nonparametric methods, each having its own advantages and drawbacks. 

Parametric methods are based on the idea that galaxy light profiles may be sufficiently well described by analytic formulas. Indeed parametric profiles have been shown to well describe galaxy structural properties since many decades now. The $R^{1 \over 4}$ law \citep{deVa48} describing almost perfectly luminous early-type galaxy light profiles, and the exponential law~\citep{free70} which instead well fits the light profile of disk-dominated galaxies, have been extensively used to derive galaxy physical quantities, such as the effective radius and surface brightness. This has in fact allowed to formulate and study galaxy scaling relations, such as the Tully-Fisher \citep{tullyfisher} or the Fundamental Plane \citep{dressler87,george87,bender1992}, which have enormously boosted our knowledge and understanding of galaxy evolution and formation. 

The $R^{1 \over 4}$ and the exponential laws may be used in the same modeling to describe both the bulge and disk photometric components of a single galaxy~\citep[see e.g.][]{simard1999}. This approach is in fact doable only for very high signal to noise objects as the number of parameters involved becomes fairly large and hence the fitting itself heavily degenerate. For this reason it has become very popular the use of the more simple and robust single component \sersic~(1968) model\footnote{We refer to Appendix A for the analytic expression of the \sersic~profile.}. This is a convenient generalization of the $R^{1 \over 4}$ profile with $R^{1 \over n}$, where $n$ is called the \sersic~index. The variation of the index $n$ allows to describe very different galaxy light profiles from the classical $R^{1 \over 4}$ ($n=4$), to the late-type galaxy exponential profile ($n=1$), to the massive elliptical galaxy profiles ($n\ge3.5$) and down the unresolved point-like Gaussian profiles ($n=0.5$). 

The \sersic~profile fitting, thanks to its flexibility and robustness,  has become a preferred tool in the last years to classify galaxies in broad morphological classes~\citep[e.g.][]{shen03,trujillo04,ravi04,barden2005,mcintosch05,P06,boris2007}, possibly combined to the ``bumpiness'' parameter as done in \citet{blak06} and  \citet{vdW08}.

The main idea behind the nonparametric approach is to measure model independent quantities, such as the
concentration, the asymmetry~(\citealt{ab1996,conselice2003}), the M$_{20}$ and the Gini
parameters~(\citealt{ab2003,lotz2004, scarlata071}) of the galaxy light distribution and to correlate these quantities to the visual galaxy appearance. Albeit being model independent, and hence conceptually more effective in describing the variegated galaxy morphology zoo, these classification schemes are not free from severe biases as addressed in \citet{blan03} and more recently in \citet{lisker2008}. 

Having in mind to continue the study presented in this paper by 
investigating also the evolution of galaxy sizes, we opted for the 
parametric approach, that delivers both morphological classifications and 
structural parameters.

In this context the COSMOS ~\citep{scov06} survey is the ultimate
effort to cover a sensitive patch of the sky ($2\sq\degr$) with HST,
going factors of 10/100/1000 wider than previous efforts like the HDF-N~\citep{HDF96}, GEMS
~\citep{rix:1}, the Great Observatories Origins Deep Survey  
(GOODS;~\citealt{giavalisco2004}), the FORS Deep Field (FDF;
\citealt{fdf:data}) among many others.

The high resolution imaging has been complemented in COSMOS by a
state-of-the-art multiwavelength coverage all the way from the X-ray
to the radio. By observing the Universe in all its colors
COSMOS is becoming a real breakthrough in observational cosmology. The need to tackle and understand the
conspiracies of Nature embedded in the Universe variance makes of this
wide and deep survey a unique and unprecedented tool for galaxy evolution studies.

In this work we rely on a deep and accurate (i $\lsim$ 26.5 and
\mbox{$\Delta z / (z_{spec}+1) \approx 0.035 $}) photometric redshift
catalog~\citep{armin2008} complemented with a morphologically classified sample (F814
$\lsim$ 24) to study the contribution of galaxies of different
morphologies to the redshift evolution of the stellar mass density, as
well as its dependence on the environmental density, over an area of
$\sim$0.7 $\sq \degr$ in the COSMOS field.

\citet{brinchmanneellis2000}, \citet{bundy2005}, \citet{P06} and \citet{fran06} 
have carried out studies on the evolution of the galaxy stellar mass function split by morphological types, 
in a way qualitatively similar to the one we present here, 
but based either on shallower samples and visual morphological classification 
or on much smaller areas heavily affected by cosmic variance. Likely due to the small surveyed area, no one of the just mentioned studies defined environmental properties of galaxies in order to contrast the effect of local environment against cosmic time on morphological evolution.

This paper is organized as follows: in \S~\ref{sec_data} we discuss
the ground-based dataset and the photometric redshifts on which this
work is based, in \S~\ref{mmll} we discuss how galaxy stellar masses
have been estimated, in \S ~4 we present the ACS data used in this
work, the source extraction and the galaxy number counts, in
\S~\ref{sec_sersic} we briefly describe the quantitative morphological
analysis, referring to a more detailed
discussion in Appendix A, in \S ~\ref{sec_mass} we present the
evolution of the morphological mass function and mass density, 
in \S~\ref{sec_env} we discuss the effect of local environments on
galaxy morphological evolution and finally in \S
~\ref{sec_conclusions} we summarize our results and draw our
conclusions.

Throughout this work, we use AB magnitudes and adopt a $\Lambda$
cosmology with \mbox{$\Omega_M=0.3$}, \mbox{$\Omega_\Lambda=0.7$}, and
\mbox{$H_0=70 \, \mathrm{km} \, \mathrm{s}^{-1} \, \mathrm{Mpc}^{-1}$}.

\begin{figure}
\begin{center}
\includegraphics[angle=0,width=0.98\linewidth,bb = 19 145 580 700]{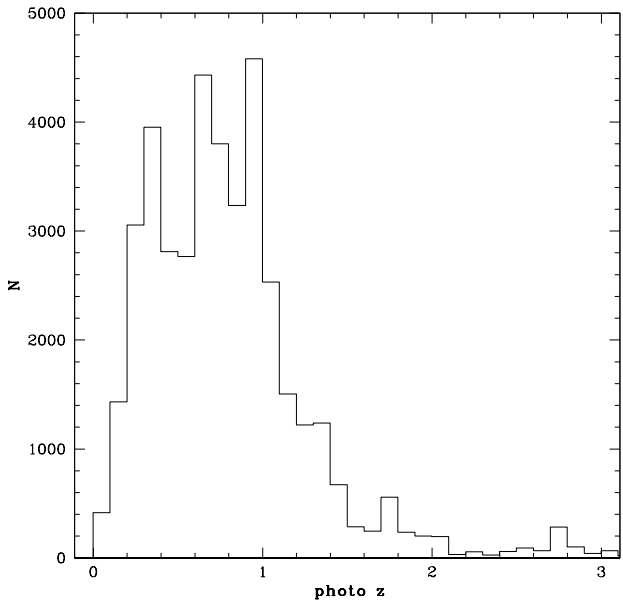}
\caption{Redshift distribution of the 41300 galaxies belonging to the
morphological catalog used in the present work. Overdensities at
redshifts 0.4, 0.7 and 1 are quite prominent, as well as the underdensities at redshifts 0.5 and 0.8.  \label{zzdist}}
\end{center}
\end{figure}

\section{Ground-based data and  photometric redshifts} 
\label{sec_data}
The ground-based data used in this paper combine the {\it uBVrizK} COSMOS dataset\setcounter{footnote}{0}
\footnote{publicly available at http://irsa.ipac.caltech.edu/data/COSMOS} with proprietary imaging in the
H band. A highly homogeneous multi-wavelength i-band selected catalog
is used to derive accurate photometric redshifts. A full description
of the NIR data acquisition, data reduction and catalog assembling,
and of the photometric redshifts estimation, is presented in
\citet{armin2008}. Here in the following, for the sake of
clarity, we give only a brief summary.

Our {\it cleaned} i-selected
catalog\footnote{publicly available at http://www.mpe.mpg.de/$\sim$gabasch/COSMOS} comprises 293~377
objects over an area of $\sim$0.7 square degrees down to a
limiting magnitude $\sim$26.5. There is a good agreement between
literature data and this catalog number counts up to the 25th
magnitude. At the bright end (i$\le$21) the literature number counts
are sensibly higher since most of the very bright objects are
saturated in the Subaru i-band images and thus not present in the
final catalog. Since objects brighter than i=21 are likely to be
important for the determination of the high--mass end of the galaxy
stellar mass function at low and intermediate redshifts, we added
those objects (1896) back to our catalog by complementing it with the COSMOS
legacy photometric catalog described in \citet{capak2007}, where CFHT
i band photometry was used for this bright tail.

Photometric redshifts were derived using the technique described in
\citet{photred00,gabasch:lf} and \citet{fabrice08}. In short, the method consists of : i)
checking photometric zeropoints and, if necessary, determining
photometric offsets by comparing theoretical and observed stellar {\it
locii}, ii) computing object fluxes in a fixed aperture ($2.0\arcsec$)
from seeing--matched images, iii) determining a redshift probability
function P(z) for each object by matching the object's fluxes against
a set of template spectra covering a wide range of stellar population
ages and star-formation histories.

The additional use of the GALEX FUV and NUV bands \citep{zamojski2007} allowed some photo-z
degeneracies to be broken, and a final accuracy of
\mbox{$\Delta z / (z_{spec}+1) \approx 0.035 $} to be reached. Figure 1 shows the redshift distribution of the subsample with available morphological information used in this work.\footnote{In the last stages of this work the COSMOS Legacy released a new catalog of spectroscopic redshifts within the zCOSMOS~DR2~\citep{lilly2007} as well as an updated catalog of photometric redshifts~\citep{ilbert2009}. We refer to Appendix B for a comparison with these two catalogs.}

The morphological catalog, including stellar mass estimates, analysed in
this work is made publicly available at the following address: http://www.aoc.nrao.edu/$\sim$mpannell/data.html.

\begin{figure}
\begin{center}
\includegraphics[angle=0,width=0.98\linewidth,bb = 19 145 580 700]{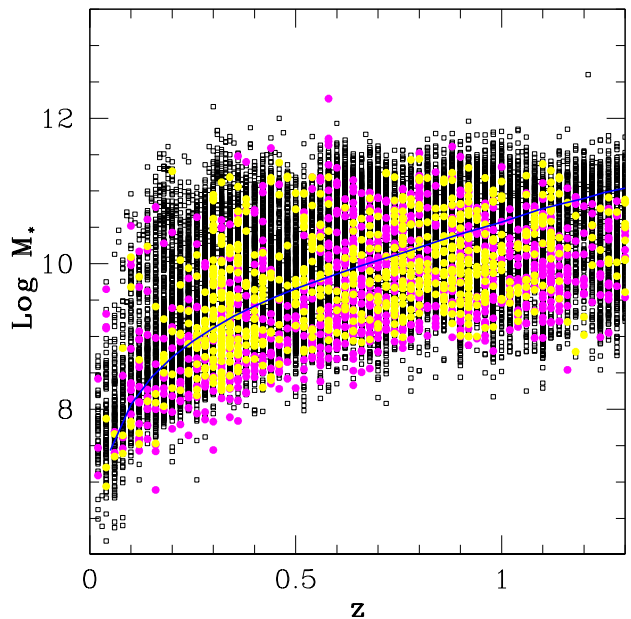}
\caption{The distribution of galaxy stellar masses, derived using a Salpeter (1955) initial mass function, as a function of
redshift for the morphological catalog used in this work (empty
squares). The solid blue line shows the estimated mass completeness of
the morphological sample at each redshift (see text for details). For
comparison, data from the morphological catalog used in \citet{P06}
are overplotted (yellow and magenta symbols for the FDF and GOODS-S
fields, respectively). \label{logmz}}
\end{center}
\end{figure}

\begin{figure}
\begin{center}
\includegraphics[angle=0,width=0.94\linewidth,bb = 19 145 580 560]{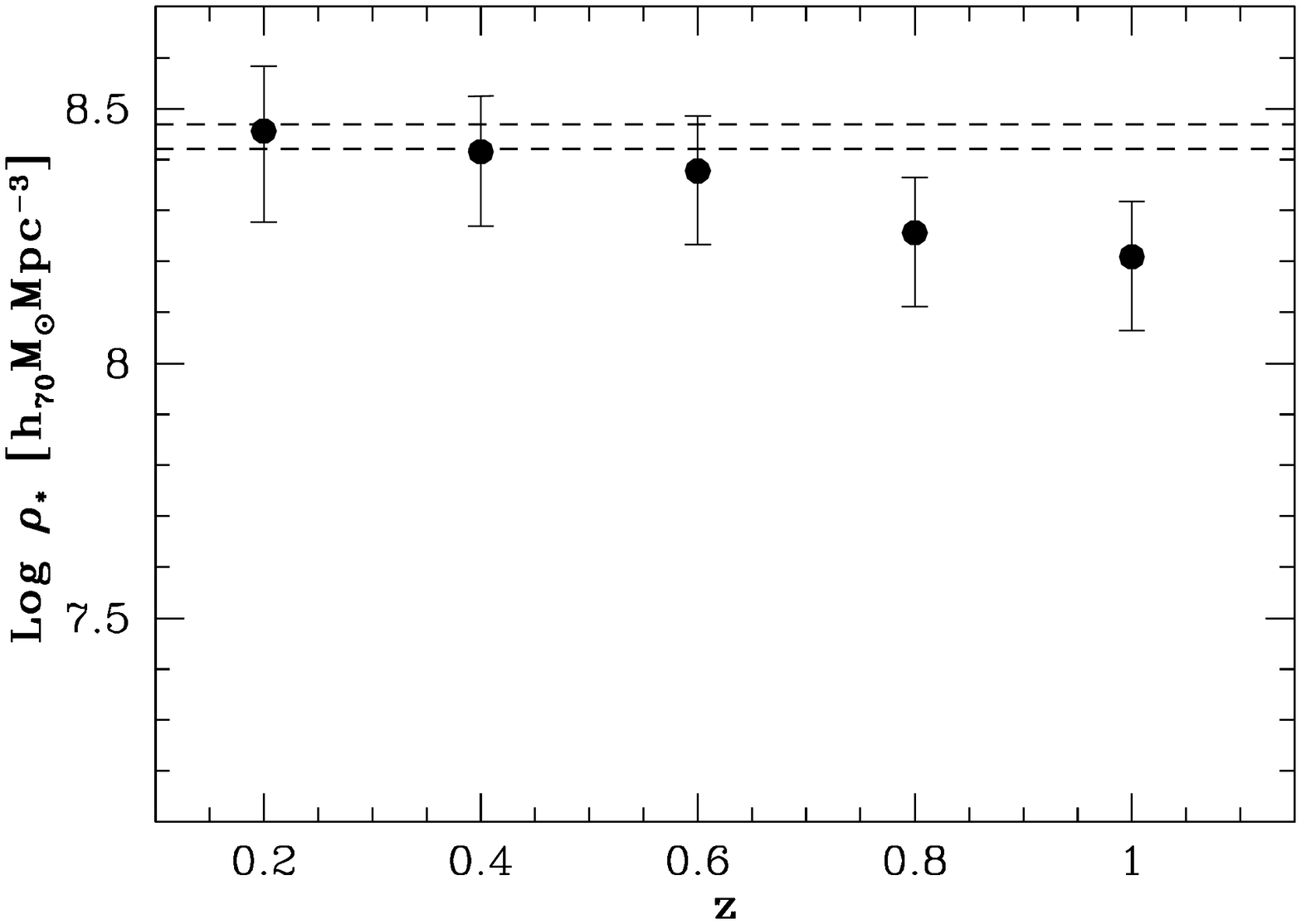}
\caption{Cosmic stellar mass density up to redshift 1 for objects with
Log~M$_* \gsim 10.8$, corresponding to the mass completeness value for
the highest redshift considered. Horizontal dashed lines show the
local mass density from \citet{bell2003} after correcting for the different IMF and applying the same mass
cut. The error bars represent the cosmic variance contribution
\citep{somerville2004}. Assuming that the mass function evolves only 
in normalization over this redshift range, adding 0.28 dex to the
plotted values gives the total mass densities.\label{rhoz}}
\end{center}
\end{figure}

\section{Computing mass--to--light ratios}
\label{mmll}

The method we use to infer stellar masses from multi--wavelength
photometry is described in detail in \citet{drory2004b}. It is based
on the comparison of the observed multi--color photometry with a grid
of stellar population synthesis models produced with the \citet{bc03}
code.

We parametrize the possible star formation histories (SFHs) by a
two-components model, consisting of a main, smooth component described
by an exponentially declining star formation rate $\psi(t) \propto
\exp(-t/\tau)$, linearly combined with a secondary burst of star 
formation. The main component timescale $\tau$~ varies ~in $\in [0.1,
\infty]$~Gyr, and its metallicity is fixed to the solar metallicity.
The age of the main component, $t$, is allowed to vary between 0.5~Gyr
and the age of the Universe at the object's redshift.

The secondary burst of star formation is modeled as a 100~Myr old
constant star formation rate episode of solar metallicity.

We adopt a \citet{salpeter1955} initial mass function for both
components, with lower and upper mass cutoffs of 0.1 and 100~\Msun.

Additionally, adopting the Calzetti et al.~(1994) extinction, both the main component and the burst are allowed to
exhibit a variable amount of attenuation by dust with $A_V^{1,2}$~$\in
[0, 1.5]$~and~$[0, 2]$ for the main component and the burst,
respectively.  This takes into account the fact that young stars are
found in dusty environments and that the star light from the galaxy as
a whole may be reddened by a (geometry dependent) different amount. We
compare this set of models with multi--color photometry of each
object, computing the full likelihood distribution in the
5-dimensional parameter space ($\tau, t, A_V^1, \beta, A_V^2$), the likelihood of each model being $\propto
\exp(-\chi^2/2)$.

To compute the likelihood distribution of mass--to--light ratios \ML,
we weight the \ML\ of each model by its likelihood and marginalize
over all parameters. The uncertainty in \ML\ is obtained from the
width of this distribution.

For young objects, with relatively high burst fractions, the width of
the \ML\ distribution is usually much wider in the optical than in NIR
bands, while for quiescent objects the width of the \ML\ distribution
is very similar in all bands. On average, the width of the likelihood
distribution of \ML\ at 68\% confidence level is between $\pm 0.1$ and
$\pm 0.2$~dex (using the B band \ML).  The uncertainty in mass has a
weak dependence on mass itself (increasing with lower $S/N$ photometry), and
it mostly depends on the spectral type: quiescent galaxies have more
tightly constrained masses than star-forming ones.

We estimated the galaxy stellar masses for all the objects in the i
band selected catalog down to the limiting magnitude.

In Figure \ref{logmz} we show the distribution of galaxy stellar
masses as a function of redshift for the morphological catalog used in
this work. For comparison, and to show the impressive statistic
available, we overplot the morphological catalogs used in
\citet{P06} from the FDF~(yellow) and GOODS-S~(magenta) fields.

In Figure \ref{rhoz} we show the stellar mass density up to redshift 1
for objects with Log~M$_* \gsim 10.8$, which corresponds to the mass
completeness value for the highest redshift considered. Dashed lines
represent 1$\sigma$ estimates of the local mass density from
\citet{bell2003} after applying the same mass cut. The error bars
represent the cosmic variance contribution, as estimated in
\citet{somerville2004}, that is a factor 10 larger than statistical
errors.  Assuming that the mass function evolves only in
normalization over this redshift range, one should add 0.28 dex to the
plotted values to recover total mass densities.

The mass completeness at different redshifts was estimated as the mass
of a maximally old stellar population\footnote{We used a dust-free,
passively evolving stellar population model, ignited by an
instantaneous burst of sub solar (Z=0.008) metallicity at z = 10.} having,
at the redshift considered, an observed magnitude equal to the
catalog completeness magnitude \citep[e.g.][]{dickinson2003}.  In
table~\ref{ttmas} we provide the mass completeness values as a
function of redshift for the morphological catalog (F814W $<$24), as
well as for the global photometric catalog (i~$< 26$).

\begin{table}
\begin{center}
\caption{Log~M$_*$/M$_\odot$ completeness versus redshift}         
\label{ttmas}      
\begin{tabular}{c c c c c c c}     
\hline\hline                
z & 0.25 & 0.45 & 0.65 & 0.85& 1.05 & 1.25\\   
\hline                       

F814W (24)& 8.9& 9.4 & 10.0 & 10.3& 10.7 & 11\\   
i$$ (26)& 8.2&  8.8   & 9.2 & 9.6&  10.0   & 10.3\\
\hline                                  
\end{tabular}
\end{center}
\end{table}

\section{HST Advanced Camera for Surveys imaging}
\label{hst}

The $2\sq\degr$ COSMOS field was imaged in the F814W ACS filter for a
total of 581 orbits. Within each orbit, four equal length exposures
of 507 sec duration each (2028 sec total) were obtained in a 4
position dither pattern, designed to shift bad pixels and to fill in
the 90 pixel gap between the two ACS CCD arrays. Adjacent pointings in
the mosaic were positioned with approximately 4\% overlap in order to
provide at least 3 exposure coverage at the edge of each pointing and
4 exposure coverage over approximately 95\% of the survey area. This
multiple exposure coverage with ACS provided excellent cosmic ray
rejection.

A full description of the ACS data processing including drizzling,
flux calibration, registration and mosaicing is provided in
\cite{koek07} and in \cite{scov06b}. 


The ACS images released to the public\footnote{Publicly available at
http://irsa.ipac.caltech.edu/data/COSMOS/} are sampled with
0.05\arcsec ~pixels. The measured FWHM of the PSF in the ACS F814W band
filter is $\approx0.11$\arcsec.

\subsection{Source extraction, star-galaxy classification and cataloging}
\label{sources}
Source extraction was performed with the SExtractor code
~\citep{bertin} on each ACS tile in a {\it cold} manner,  in order
to minimize the artificial source splitting and maximize the number of
pixels assigned to each object.

We used 348 ACS tiles, namely only those overlapping with the ground
based photometry used in this work.

Taking advantage of the ACS high resolution imaging we could remove
the point--like (stellar) objects using the structural parameters
output by SExtractor. We used the full width half maximum (FWHM), the
half light radius, the neural network stellarity index and the total
magnitude for each object to select point--like sources.

\begin{figure}
\begin{center}
\includegraphics[angle=0,width=0.94\linewidth,bb = 19 145 580 700]{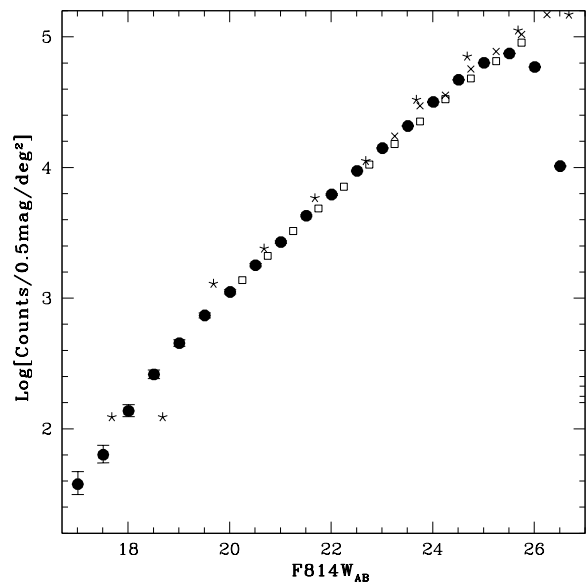}
\caption{Galaxy number counts for the ACS/F814W passband are plotted as
solid dots. Error bars account for poissonian errors only, and no
correction for incompleteness has been applied. Literature data are
plotted with other symbols and are taken from: \citet{P06}~(asterisks, FDF), \citet{leat07}~(open squares, COSMOS) and \citet{ben}~(diagonal crosses, VV~29). \label{nncc}}
\end{center}
\end{figure}

In Figure~\ref{nncc} we show F814W band galaxy number counts, not
corrected for incompleteness, compared with literature counts~\citep{P06,leat07,ben},
and in table~\ref{tab:nc} we provide the plotted values. The ACS
galaxy catalog has been cross-correlated with the ground based catalog
with a 1 arcsec matching radius. We removed from the final catalog all
the objects, a few percent of the total, for which one ground based entry had 
more than one entry in the ACS catalog. The final
morphological catalog used in the following contains 41300
objects over $\sim$0.7 $\sq \degr$ and results to be more than 90 \%
complete down to F814W = 24.

\begin{table}

\caption{Galaxy number counts not corrected for incompleteness in the
$F814W$ passband ($\sim$0.7 $\sq \degr$). $\log N$ and $\sigma_{\log N}$ are 
given, where $N$ is in units of $\mathrm{0.5 mag}^{-1} \mathrm{deg}^{-2}$.}
\label{tab:nc}
\begin{center}
\begin{tabular}{ccc}
\hline
\hline
& $F814W$ &   \\
$mag$ & $\log N$ & $\sigma_{\log N}$   \\\hline
17.01 &    1.576  &   0.0969 \\ 
17.51 &    1.801  &   0.0728 \\  
18.01 &    2.137  &   0.0481 \\  
18.51 &    2.416  &   0.0343 \\  
19.01 &    2.655  &   0.0258 \\  
19.51 &    2.867  &   0.0201 \\  
20.01 &    3.047  &   0.0163 \\  
20.51 &    3.252  &   0.0128 \\  
21.01 &    3.429  &   0.0104 \\ 
21.51 &     3.63  &   0.0082 \\  
22.01 &    3.793  &   0.0068 \\  
22.51 &    3.974  &   0.0055 \\  
23.01 &    4.148  &   0.0045 \\  
23.51 &    4.318  &   0.0037 \\  
24.01 &    4.501  &   0.0030 \\  
24.51 &    4.671  &   0.0025 \\
25.01 &    4.802  &   0.0021 \\ 
25.51 &    4.872  &   0.0020 \\  
26.01 &    4.769  &   0.0022 \\  
26.51 &     4.01  &   0.0053 \\ 
\hline
\end{tabular}
\end{center}
\end{table}

\section{The morphological analysis}
\label{sec_sersic}

We use the package GIM2D ~\citep{simard1999} to fit PSF
convolved \citet{sersic1968} profiles to the two-dimensional surface
brightness distribution of each object, for all sources down to a
limit of F814W=24. The PSFs used to convolve the profiles were
obtained for each individual tile by median stacking about 50 high S/N
stars.

We tested our results by running also the GALFIT ~\citep{peng2002}
code on some of the COSMOS ACS patches. The results from the two
different codes are in excellent agreement, thus confirming both the
robustness of our modelling and the choice of flux limits.

We further tested our results by running the GIM2D code on simulated
images. This test was performed by adding to the real image, in a
blank, pure--sky region, fake objects with \sersic~profiles and
structural parameters spanning a wide range of plausible values.  The
whole procedure is then run exactly as for real objects: we find
that the profile parameters of simulated objects are well recovered
down to F814W=24. High \sersic~index objects have very extended wings
that, depending on the total flux, can fall under the sky surface
brightness, thus for these objects a lower \sersic~index, total flux
and effective radii are usually recovered (see also \citealt{sargent2007}). We point out that the same
trend was found in \citet{tru2006} with a different fitting code,
thus confirming the "physical bias" as contrasted to a code
failure. In the local Universe a tight correlation between \sersic~index and luminosity/mass of giant elliptical galaxies ~(see e.g.~\citealt{caon1993} and \citealt{graham2001}) has been robustly established. Due to flux limits and mass completeness, morphological analysis of high redshift ellipticals is mostly limited to the most massive galaxies which one would expect to have relatively high \sersic~index.
Our  simulations show that for such high \sersic~index objects at faint apparent magnitudes (low S/N) we would recover \sersic~indexes and effective radii biased toward lower values. While we will deal in detail with the galaxy size evolution in a future work, we argue here that the strong size evolution of early-type galaxies with redshift, recently reported in the literature \citep[e.g.][]{tru06,cim2008,vd2008}, could be partly due to such bias. We try to cope with this "biased" \sersic~index by adopting the  procedure described in the following. We perform extensive Monte-Carlo simulations to take into account the \sersic~index uncertainty ($\approx$30\%, estimated through simulations) on our results. One thousand simulations of the morphological catalog were generated. Objects with \sersic~index smaller than 2.5 were perturbed within a Gaussian of sigma equal to the 30\% of the measured value. Objects with  \sersic~index larger than 2.5 were perturbed within only the positive side of the Gaussian so that their value could only get larger than the measured ones. We tried also to implement different solutions to this correction procedure and the final results were always in good agreement.   
More details on the morphological analysis are given in the Appendix A.

\section{The evolution of the galaxy stellar mass function by different morphologies}
\label{sec_mass}

\citet{P06} show that there is a good correlation between the average visual and automated classifications as parametrized by the morphological type $T$, assigned according to the \citet{devauc} classification scheme, and the \sersic~index $n_{ser}$, respectively (see Figure~1 in that paper). Following \citet{P06}, we split our sample, according to the \sersic~index value in three broad morphological classes: early-type (n $\ge$ 3.5, which corresponds to bulge-dominated/elliptical galaxies), intermediate (2 $\le$ n $<$ 3.5) and late-type (n $<$ 2, which corresponds to disk-dominated galaxies).

We restrict our analysis to $z \le 1.2$ to limit the bias introduced
by the rest-frame emission band shifting, namely the effect of the
morphological k-correction. A number of studies in the last years have
shown, although not in a conclusive and quantitative way, that this is
indeed a fair assumption \citep[e.g.][]{scarl06, boba07}.

\begin{figure}[htbp!]
\begin{center}
\includegraphics[angle=0,width=0.48\textwidth,bb = 110 80 400 720,clip]{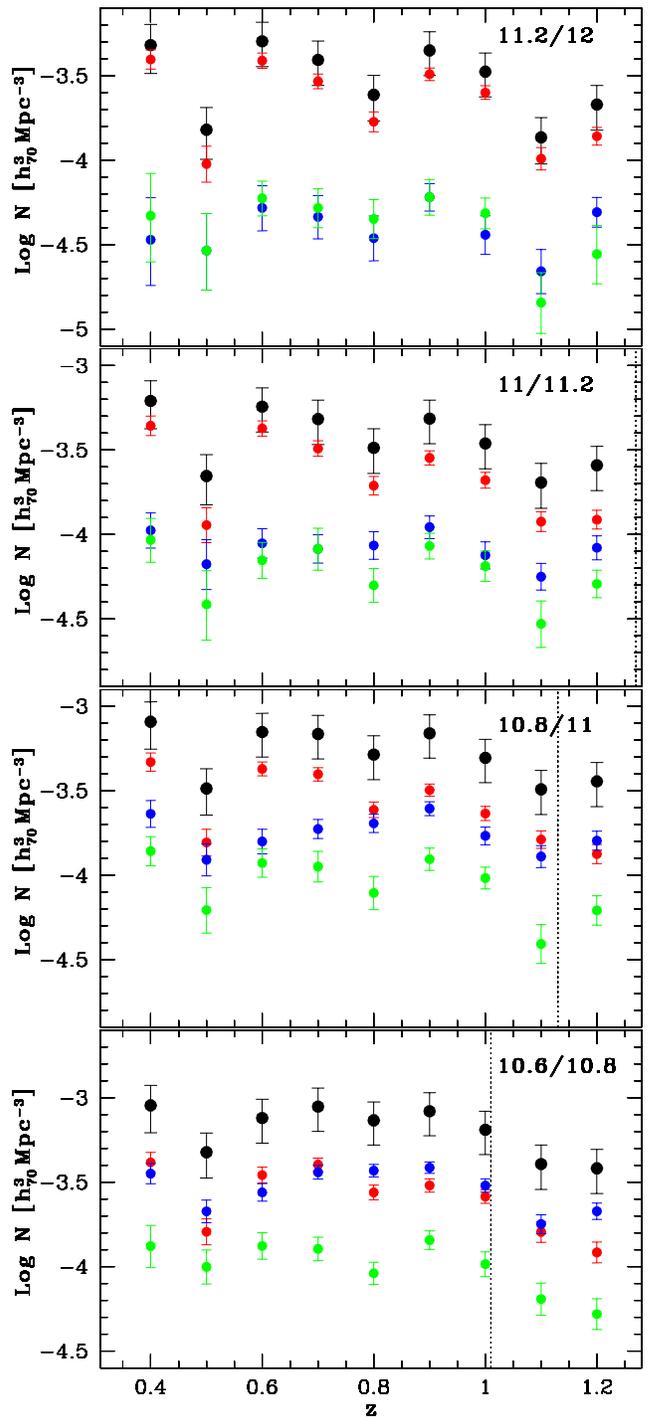}
\caption{Number densities as a function of redshift and
morphology. The panels show the number density for different stellar
mass bins, as indicated in units of Log(stellar mass) in the upper
right corner of each panel. Vertical dotted lines show the limiting
redshift for which the sample in the considered stellar mass bin is
complete. 
Symbols are colored according to the morphological split (black-total,
red-early, green-intermediate, blue-late).}

 \label{nndd}
\end{center} 
\end{figure}

We performed extensive Monte-Carlo simulations to take into account
the effect of mass uncertainties ($\approx0.2$dex) on our results. One
thousand simulations of the mass catalog were generated, perturbing
each mass within a Gaussian of sigma equal to its error. Unless stated
differently in the relevant Figure captions, we use the median values
of the Monte-Carlo simulations in all Figures. Error bars take into
account both poissonian errors on the median counts ~\citep{gehr1986},
and 16-84$^{th}$ percentile values of each distribution.

In Figure~\ref{nndd} we show the V/V$_{max}$ corrected number density
evolution split by morphological type, up to redshift 1.2, for objects
in stellar mass ranges Log~M$_*$=[10.6-10.8], [10.8-11],
[11-11.2], and [11.2-12], as indicated in the upper right
corner of each panel. By multiplying the number densities by
the average mass in the corresponding mass range, each panel can be
interpreted as the evolution of the total mass densities contributed
by objects in that stellar mass range. The dotted vertical lines show
the estimated redshift completeness for the relevant mass bin.

We find that early--type objects always dominate the high--mass end. 
The lower the stellar mass, the more late--type objects tend to
dominate the relative contribution to the total number/mass
density. The same trend is true by looking at increasing redshift: the
higher the redshift the more the relative contribution of early and
late types gets closer.  For the lowest mass range [10.6-10.8] the
transition from a bulge--dominated to a disk--dominated stellar mass
budget happens around redshift 0.7; beyond that redshift late-types are
always contributing more to the total mass budget. In the mass range
Log~M$_*$ = [10.8-11] the transition is occurring at $z \approx 1.1$.

At each redshift, a {\it transition mass} can be identified where the
transition from a bulge--dominated to a disk--dominated stellar mass
budget takes place.  Based on the very good statistics available in
this work, we are able to confirm the main conclusions of Pannella et al. (2006): {\it
i)} the morphological mix at the high--mass end evolves with redshift;
{\it ii)} the transition mass increases with redshift. At $z\sim 0.7$
we find a value of the transition mass approximately consistent with
the local value $M\approx5\times10^{10}$M$_\odot$, as measured in
\citet{bell2003} by using the concentration parameter to discriminate
between early and late type objects in a complete sample extracted
from local surveys. At $z\sim1$ the late and early types contributions
become comparable at $M\approx1\times10^{11}$M$_\odot$.

In a fixed mass range the number densities of late-types are consistent
with being constant with redshift, while the early types follow the
more general total declining. 
At a fixed redshift, the number density of early types of different
masses is very similar (in +- 0.1/0.2 dex), while the late type number
densities move by one order of magnitude passing from the high to the
low stellar mass range. This naturally sets the evolution of the
transition mass with redshift.

The slope of the redshift evolution of total number densities is
consistent in all mass ranges, at least up to the common redshift
completeness, meaning that the mass function in its high-mass end, i.e. at least down to  $M\approx4\times10^{10}$M$_\odot$, does
not evolve in shape but only in normalization with redshift.

\section{Disentangling the role of environment on\\ galaxy morphological evolution}
\label{sec_env}

In Figure~\ref{nndd} there is obvious scatter around the mean number
density decline in all the mass bins. However, an underdensity at z$\sim$0.5 shows up quite strikingly. 
The deficit of objects is also quite
evident in the global redshift distribution of our morphological
catalog, as shown in Figure~\ref{zzdist}. It gets more and more
significant with increasing stellar mass. The deficit of galaxies at
this redshift is a factor of three for the highest mass objects, and
less than a factor of two for objects in the lowest mass bin
explored. This differential deficit is in qualitative agreement with
galaxy evolution dependence on the local density as expected in
current galaxy formation models (see \citealt{lee06} and references
therein): the more underdense is the environment the later
massive objects will assemble because there is in general a time delay in
starting star formation and also there are less small galaxy units to
assemble the massive giant galaxies.

From the morphological point of view, the evolution in this underdense
redshift bin is {\it delayed} at all stellar masses as compared with
the contiguous redshift bins. The early-type fraction with respect to both
the total number density and the late-type number density is much
lower than those in the contiguous redshift bins, in agreement with
the expectations from the morphology-density relation. We also note that
in the highest mass bin, in this underdense region there are no
massive late-type galaxies.

We notice that the morphological mix in this underdense redshift bin
is, at all masses probed, very close to the morphological mix at z$\approx$1.1, which has a comparable total number density of objects.

Taking advantage of the good statistics available in this study, we
wish to further check the importance and role of environment on
galaxy morphological evolution.

Estimating the local volume density in a photometric redshift survey is
not straightforward, and we try here to make it as simple and
unbiased as possible. 

First of all one has to decide which are the objects tracing the
overdensities in a way not biased by the flux limited
survey: a density-defining population (DDP, \citealt{croton2005}) of galaxies 
is needed. The objects have to be bright enough to be seen in
the whole redshift range considered. Dealing with masses and
morphologies, one should also take care of the selection effects
introduced by the different M/L ratios associated with different
stellar populations. Basically, old stellar populations, which more
often but not always reside in bulges, have much higher M/L ratios
with respect to young stellar populations, often but not always
populating disks. A pure luminosity cut -- even taking into account
passive evolution -- would thus preferentially cut out of the sample
high M/L objects (early-types) with respect to low M/L objects (late-types). To
avoid this kind of bias, we use as DDP all objects
with masses down to the mass completeness value at the highest
redshift we are exploring (7x10$^{10}$ solar masses).  

Another issue is the relative scarcity of these objects (being in the
exponential cutoff of the mass function), so one has to
avoid reaching the shot noise regime when calculating number
densities.

With this in mind, we proceed as follows: we first fit the number
density evolution with redshift for objects more massive than
7x10$^{10}$ solar masses, obtaining in this way an expected value of
the number density,~$\bar{\rho}(z)$, at each redshift z. We then split
the entire redshift range in slices of $\Delta z = 0.1$ (corresponding
to differential depths of about about 327 comoving Mpc at redshift 0.5
and 244 at redshift 1). This slicing along the redshift axis turns out
to be quite robust against the photo-z errors, in fact this
corresponds to about 2 times the formal photo-z error at redshift
0.5 and 1.5 times at redshift 1. We then split the entire survey volume in cells of 2.8 comoving
Mpc by side at each redshift slice, that is the maximum contiguous
angular area we have available in the lowest redshift bin. Finally, we
calculate the comoving number density of each cell at the different
redshifts and assign the cell galaxy population to one of three
density classes: underdense ($\rho(z) \le 0.5\times \bar{\rho}(z)$),
medium-dense ($0.5\times\bar{\rho}(z) < \rho(z) \le
1.4\times\bar{\rho}(z)$) and overdense($\rho(z) >
1.4\times\bar{\rho}(z)$). The limits used here to define environmental classes are chosen to balance the needs to map sensibly different environments and to assign a relatively large number of objects to each of the three classes.

\begin{figure}
\begin{center}
\includegraphics[angle=0,width=0.94\linewidth,bb = 25 145 580 510]{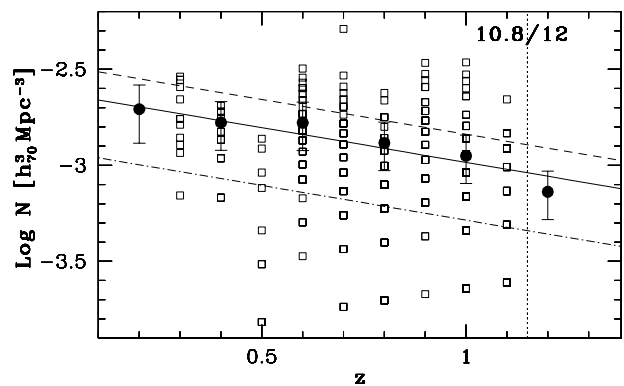}
\caption{Number density evolution for objects with Log~M$_* \gsim$ 10.8. The solid line, $\bar{\rho}(z) = -2.62 -0.36 \times z$, shows the linear fit to the filled dots (i.e. the number densities obtained over the whole field in redshift bins of $\Delta z$ = 0.2). The dashed line sets the 1.4$\times \bar{\rho}(z)$ limit above which we select the overdense regions. Below the dashed-dot (0.5$\times \bar{\rho}(z)$) line we select the underdense regions, while in between the two lines we select the medium-dense regions. The empty squares overplotted show the number densities in all the cells we have split our survey volume.\label{selct}}
\end{center}
\end{figure}

\begin{figure*}
\begin{center}
\includegraphics[angle=0,width=0.85\linewidth,bb = 19 145 580 700]{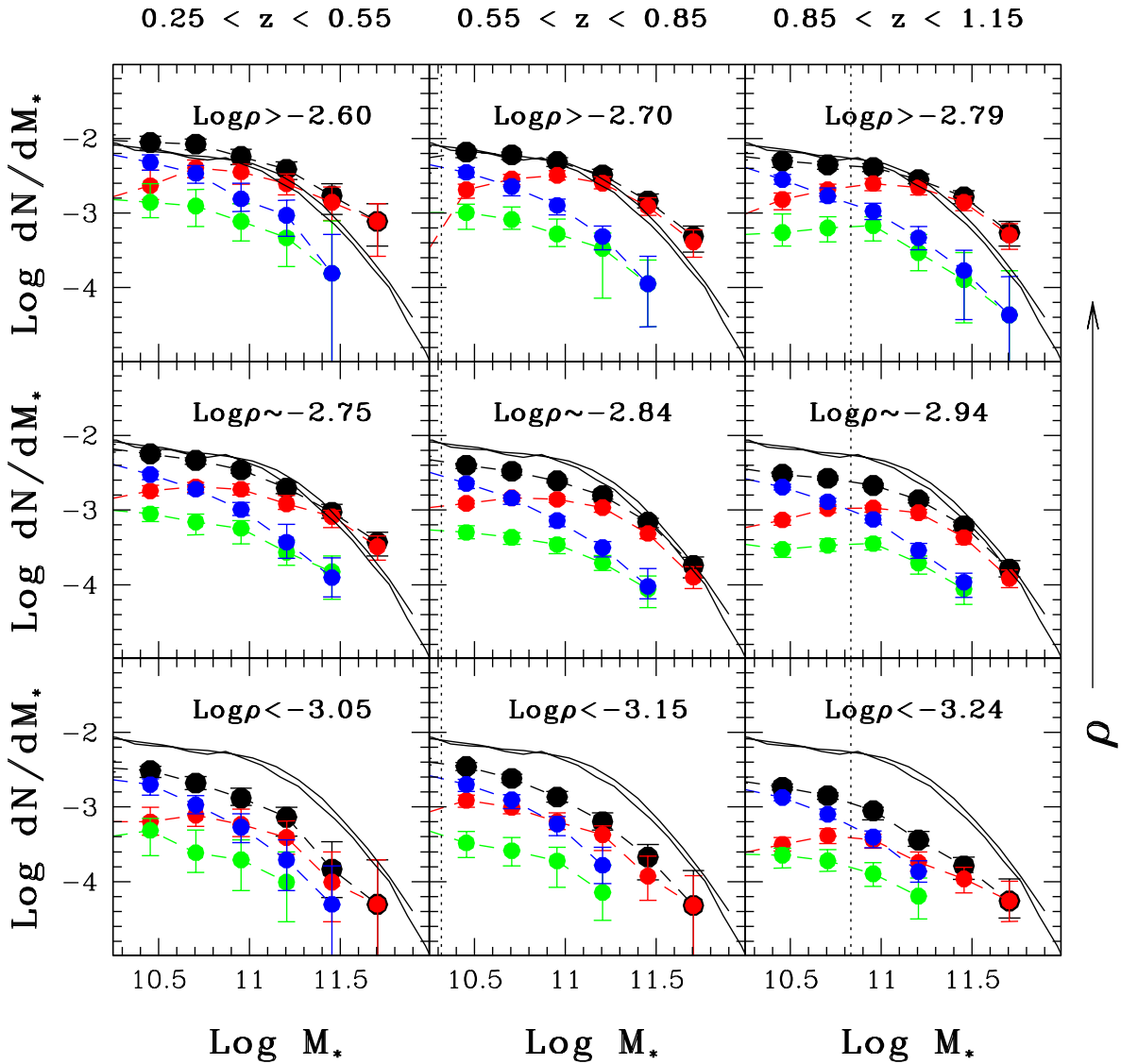}
\caption{Galaxy stellar mass functions as a function of redshift, morphology and local environment. Black symbols refer to total values. The vertical axis is in units of [h$_{70}^3$Mpc$^{-3}$dex$^{-1}$]. A vertical dotted line indicates the mass completeness limit in each redshift bin. Colored symbols refer to morphological classes, see Figure 5 for details. Dashed colored lines are only intended to guide the eye. Solid black lines show the local {\it filed} mass function determinations from \citet{cole2001} and  \citet{bell2003}.\label{mf_env}}
\end{center}
\end{figure*}

In Figure \ref{selct} we show the number density evolution for objects
with Log~M$_* \gsim$ 10.8. The solid line, $\bar{\rho}(z) = -2.62
-0.36 \times z$, shows the linear fit to the filled dots (i.e. the
number densities obtained over the whole field in redshift bins of
$\Delta z$ = 0.2). The dashed line sets the 1.4$\times \bar{\rho}(z)$
limit above which we select the overdense regions. Below the
dashed-dot line ($0.5\times\bar{\rho}(z)$) we select the underdense
regions, while in between the two lines we select the medium-dense
regions. The empty squares overplotted show the number densities in
all the cells we have split our survey volume. 

We run 100 realizations of the procedure adopted to define environmental densities by randomizing the initial position of the grid in the sky and also by adopting different solutions for the redshift slicing of the total volume. The outcome of this testing convinced us that the results we are going to show and discuss in the following are robust and not biased against the specific procedure.

\subsection{Morphological evolution and mass growth of\\ early and late-type galaxies in $0.25 < z < 1.15$}
\label{8gyrs}

In Figure~\ref{mf_env} we show the galaxy stellar mass function as a
function of redshift, split by morphology and for the different
environmental densities as defined in the previous section. Each of the
panels contains about 2000 objects. Vertical dotted lines set the
estimated mass completeness at the different redshifts. For reference
in all panels are overplotted local estimates for the stellar mass
function~\citep{cole2001,bell2003}, which are fixed and independent from the environment. They
are only meant as a reference to look at differential evolution in
different environments.

In the lowest range explored, at $z\sim0.4$, the mass completeness
reaches masses as low as 7$\times$10$^{9}$M$_\odot$. This enables us to explore the behavior of
the slope of the mass function in different environments. The slope of the mass function is marginally evolving from the lower to the higher density regions, becoming increasingly shallower and consistent
with the local {\it field} one. This trend is in fact in good agreement with what
found in the local Universe ~\citep{baldry2006}.

The transition mass moves from Log~M$_* \simeq 10.6$ at high density to
Log~M$_* \simeq 10.9$ at low density.  

In the overdense and medium-dense environments the shape of the mass
function is in a very good agreement with the local determinations. In
the overdense regions, this happening at all redshifts, a very massive bulge-dominated population seem to appear that overcomes the local {\it
field} determination.

\begin{figure}
\begin{center}
\includegraphics[angle=0,width=0.99\linewidth,bb = 19 145 580 480]{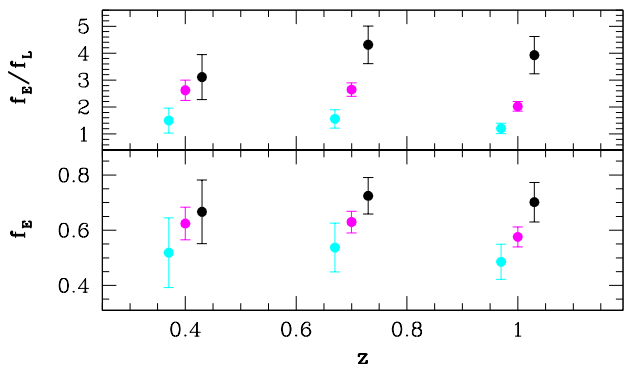}
\caption{The morphology-density relation at work. The bottom panel shows the
early-type fraction as function of redshift and environment. The environment
is color coded as black-high, magenta-medium, cyan-low. The upper panel shows
the early-to-late type classification ratio for different redshift and different
environments. All the fractions are computed for objects more massive than Log~M$_* >$ 10.8 \label{mdr}}
\end{center}
\end{figure}

\begin{figure*}
\begin{center}
\includegraphics[angle=0,width=0.9\linewidth,bb = 19 450 580 710]{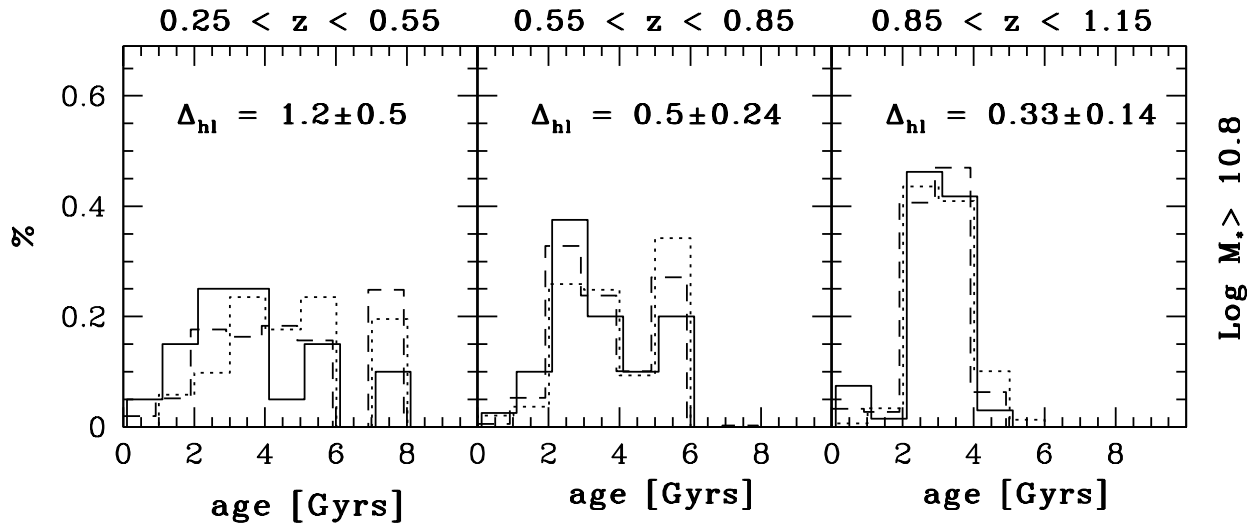}
\caption{Ages of massive ($\ge$ 7$\times$10$^{10}$M$_{\odot}$) early--type objects at different redshift and for different environments. 
The difference, in billion years, between the mean ages of early--type galaxies in high
and low density environment is indicated in each panel. The dotted,
dashed and solid histograms represent the high, intermediate and low
density environments, respectively.}
\label{ages}
\end{center}
\end{figure*}

Figure~\ref{mf_env} shows that the morphologically split mass function
evolves along both the redshift and the density axes. The evolution is
differential though: the redshift evolution is much milder than the
environmental one.

This is in fact a crucial point in interpreting the results since in
almost all previous studies there has been no chance to split these
two very different contributions.  This is better explained by the
following examples: the total mass function, and the morphological
mixtures are basically the same, up to the common mass completeness
value, for overdense volumes at redshift 1 and medium-dense volumes at
redshift 0.4. The transition mass is close to Log~M$_*$ = 11 at redshift 1
in medium-dense volumes and it is already at that mass in the
underdense environments at $z\sim0.4$. This shows how by analyzing
small patches of the sky one could mix up different environments
at different redshifts and hence get the wrong evolutionary paths for
different morphologies.

With the aim of trying to better quantify the morphological mix
evolution as a function of redshift and environment, we integrate the
mass function over the common complete mass range in the three
redshift bins and in the three different environments.

In Figure~\ref{mdr} we show both the early-type fraction (lower panel) and
the early-to-late type classification ratio (upper panel) at different redshifts and
for different environments (low--cyan, medium--magenta, black--high density).

At all redshifts a morphology-density relation, as already reported in the literature \citep[e.g.][]{smith2005}, is showing up: denser volumes contain also a higher relative fraction of bulge-dominated galaxies with
respect to lower density volumes. Although it is difficult to compare with previous results given the fact that in our study for the first time, to our knowledge, a mass complete sample has been adopted, it is very reassuring that qualitatively we are in agreement with previous studies. This confirms the robustness of our galaxy environment definition procedure.

As a function of redshift, in the medium and low density environments
there is an increase of the early-type population with cosmic time which
is more pronounced between redshift 1 and 0.7 and then flattens out to
redshift 0.4. In the high density environments the early-type fraction
grows between redshift 1 and 0.7, and then seems to turn-over down to
redshift 0.4. A similar behavior has been actually found in \citet{capak2008}.  
In our opinion the sky patch, we are using here, might not cover
enough cosmic volume at redshift around 0.4 to allow an exact
environmental definition. This shows up in Figure~6 where all but two
sub--volumes are in fact assigned to the medium dense environment bin.

\subsection{The ages of early-types in different environments}
\label{ages}
In Figure~\ref{ages} we show the ages of  massive early-type
galaxies as a function of redshift and for different environments as
obtained from the SED fitting procedure outlined above. The redshift ranges considered are indicated in the top label of each panel. 
We cut the
total sample to the common completeness limit ($\ge$
7$\times$10$^{10}$M$_{\odot}$). The dotted, dashed and solid
histograms represent the high, medium and low density environments,
respectively.  In each panel, the difference in billion years between the mean ages in
the two most extreme environments (the high and low density regions)
is indicated. Such differences deviate from zero at a two sigma level
at all redshifts, and might suggest an earlier formation epoch of
early-type galaxies in high density environments with respect to those
living in a low density region. The results found are in good agreement
with recent studies \citep[e.g.][]{thomas2005,vD2007,gobat2008,rettura2008}.

From Figure~\ref{ages} three things are worth noting: first, the
histograms are quite similar to each other within each single panel~(as confirmed by a Kolmogorov-Smirnov test) but for the oldest and youngest tails of the distribution which preferentially belong to the highest and the lowest density regions; second, the width of the distributions is increasing with the cosmic clock, this being partially driven by the narrower range of ages allowed at earlier cosmic times; third, in the lowest redshift bin a number of, relatively young, early-type objects are present which cannot be the descendants of the higher redshifts early-type populations simply evolved by passive evolution.

We interpret these evidences as follows: i) young massive
early--type galaxies are preferentially found in low density
environments at all redshifts, ii) the bulk of the early--type stellar populations have similar characteristic ages independent of the environment; iii) the oldest objects in the Universe, or better saying the oldest stars in the Universe, happen to belong to its highest density
regions; iv) newly formed, or somehow rejuvenated, bulge-dominated galaxies
are entering the sample with cosmic time. This latter finding might explain why the difference in age seems to increase with the cosmic clock and reach the 2~Gyrs quoted by Thomas et al. (2005) in the local Universe.

The estimate of stellar population ages suffers from some important
degeneracies and has to be treated with {\it a pinch of salt}. The SED
modelling we have described in section 3 only considers stellar
population models with solar metallicity. This means we are not
tackling the well known age-metallicity degeneracy: old stellar
populations with low metallicities and younger stellar populations
with high metallicities may well have very similar SEDs. Only recently \citet{cooper2008}, but see also \citet{mouchine2007} for different conclusions, reported on a mild 
observed correlation between metallicity and environment: massive galaxies in high-density environments are slightly more metal rich than galaxies belonging to low-density environments. If such a correlation is indeed at work our results are biased and the correct differences in age would be even smaller than the ones reported in the panels of Figure~9.

We remind that our classification scheme does not allow any separation between ellipticals  and early-type spirals. Therefore, as happens in the local Universe~\citep[see e.g.][]{dressler1980} our early-type samples will be likely dominated by elliptical and S0 galaxies in overdense environments and by early-type spirals in low density regions.

\subsection{Red sequence and blue cloud}
\label{sec_ubcoll}
\begin{figure*}
\begin{center}
\includegraphics[angle=0,width=0.85\linewidth,bb = 19 145 580 710]{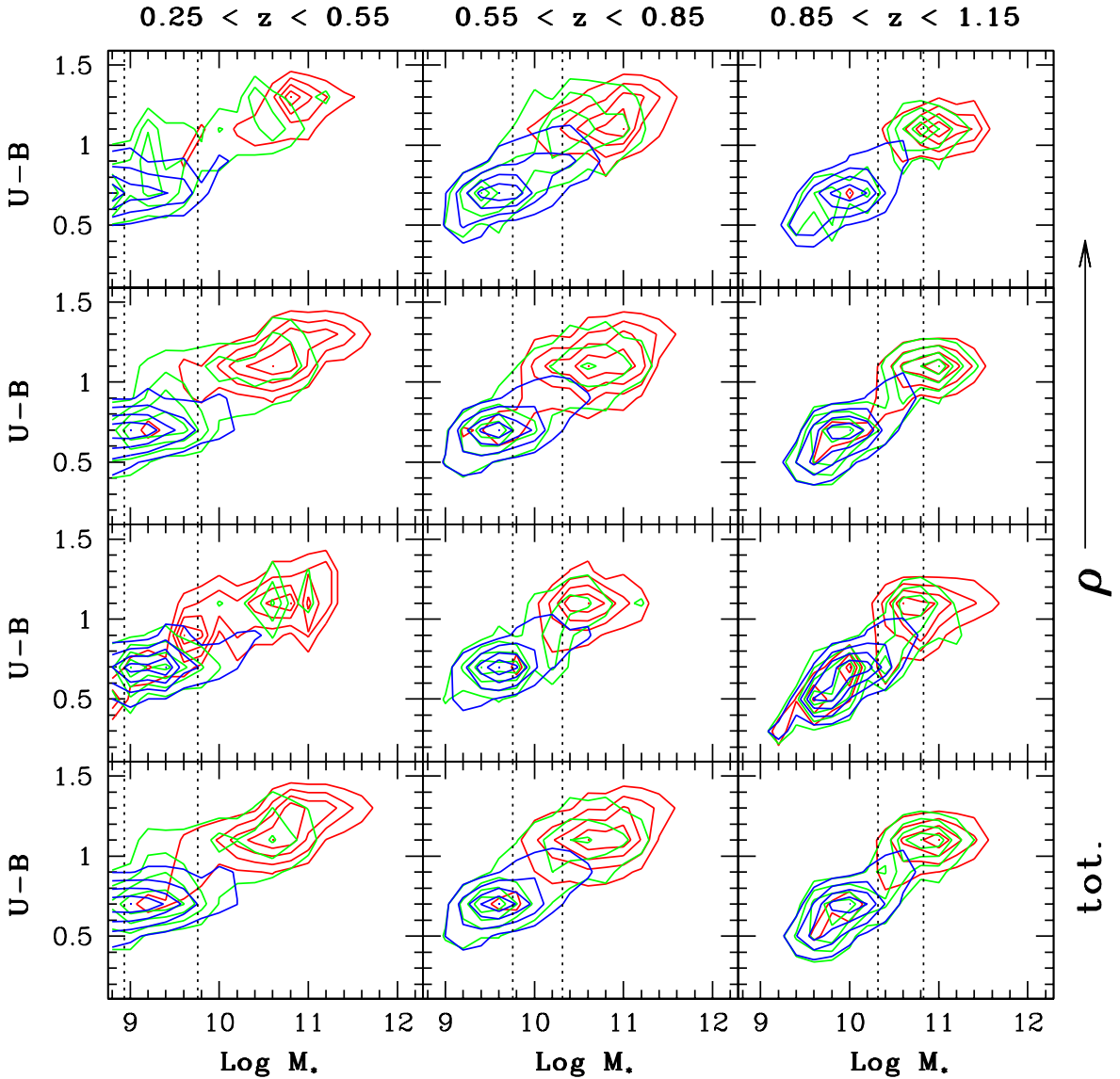}
\caption{The U-B rest-frame color as a function of mass, redshift, morphology and environment. The color coding is according to the morphological classes (red--early, green--intermediate, blue--late). The bottom row shows the total values, i.e. the first three rows, from the top, collapsed. Contours define the morphological classes and are drawn as explained in the text. The two dotted lines show the estimated mass completeness values of our sample at the upper and lower edges of each redshift bin.}
\label{ubcoll}
\end{center}
\end{figure*}

In Figure~\ref{ubcoll}~we show the rest-frame U-B color as a function
of mass, redshift, morphology and environment. Morphological
classification is highlighted in different colors (red = early, green
= intermediate, blue = late).

Contours are used to better show the different behavior of the three
morphological classes. The contours correspond to equally spaced
levels between the minimum and maximum density of the data points of
each morphological set. The first three rows of panels starting from
the top, show the high, medium and low density environments.  The
bottom row shows instead the total population, {\it i.e.} all the
different environments together. Redshift ranges are indicated, for
each column, in the top label.

The red-sequence, mostly populated by early-types, in all environments
becomes slightly redder with cosmic time because of the ageing of the
stellar populations. The evolution in the rest-frame U-B color is
mild, as also shown for instance by ~\citet{willmer2006} up to
z$\approx$1.3. The red-sequence upper envelope reddens by $\Delta
(U-B) \approx 0.25$~mag in the redshift range 0.25$< z <$1.15, in
agreement with simple passive evolution expectations.

The bulge-dominated galaxy population is mostly confined to the
red-sequence with some peculiarities dependent on environments. At all
redshifts, in higher density environments early-types tend to be more
confined to the massive end of the red-sequence.  On the other hand,
in lower density environments early-types tend to populate the red-sequence
down to lower masses (as it is easier to see at low redshifts) and
reach into the blue-cloud. This definitely points toward a difference
in age between the bulge-dominated populations in different
environments.

It is worth noting the evolution of the intermediate--type population:
these are mainly spiral galaxies hosting fairly massive central
bulges. At redshift 1, intermediate--type galaxies overlap in the
color--mass diagram with both the early and late populations in all
environments. As time goes by, this class of objects tend to leave the
very high mass end of the red sequence in favour of bulge-dominated
galaxies, and to drift toward the bluer late-type population. This suggests
that part of these intermediate--type galaxies have moved to the early-type
population. The change of this class of objects with cosmic time
correlates with the environmental density: 
at low redshifts, intermediate--type objects in more dense
environments tend to occupy the {\it valley} between the red--sequence
and the blue--cloud, while in low density environments they split in
two populations, with the majority well overlapping the late-type
population, and a minor sub--population remaining in red-sequence
together with the massive bulge-dominated galaxies.

The late-type population is confined at all redshifts and in all
environments to the blue cloud. Given the mass completess limits it is
definitely hard to make strong conclusions, but these data suggest the
presence, possibly more pronounced at higher redshift and in lower
density environments, of a population of massive disk-dominated galaxies
reaching toward the red-sequence.

In section 7.1 we have investigated the evolution with redshift of the
morphology--density relation~(see Figure~8).  Since Figure~10 shows
that at all redshifts early-types populate mainly the red sequence while
late-type galaxies are mostly in the blue cloud, we can infer from our data that a
color--density relation is in place up to the highest redshift
explored.

\citet{cucciati2006} analysed a sample of 2900 galaxies with secure
spectroscopic redshift determinations drawn from the VIMOS--VLT Deep
Survey~(VVDS, \citealt{lefevre05}) in an area 4 times smaller than the one
considered in this work and with $\approx$32\% sampling, slightly
biased toward faint objects, of the whole galaxy population. They
obtain a very similar result to the one we present in this work up to
redshift 1.2, while claiming a dramatic evolution in the redshift bin
1.2--1.5 with blue star--forming galaxies belonging preferentially to
the highest density environments. Since they define environmental
properties based on a volume limited sample, i.e. an absolute
magnitude cut at all redshifts, which is biased against old stellar
populations, it would be interesting to check whether this effect
persists if defining environmental densities with a mass complete sample.
We postpone a more detailed investigation of this issue to a forthcoming paper.
As of today, only a photometric reshift survey like COSMOS, because of
the high statistic available, may in fact allow such an investigation.

\citet{cassata2007} analysed a sample of 2041 galaxies with
photometric redshifts within 0.61 $<$ z $<$ 0.85 and covering an area
of 270 arcmin$^2$, that is about 9 times smaller than the one we use
here. The accuracy of the their photometric redshifts is comparable to
the one in this work (see \citealt{mobasher2007}). They perform a
morphological classification up to I$_{AB}$ $<$ 24, thus the same flux
limit we use in our analysis, using a set of nonparametric
measurements of the galaxy light distribution
\citep{conselice2003,ab2003,lotz2004}. Also, following
\citet{dressler1980} they estimate galaxy projected densities.  They
do find a well defined morphology-density relation in place and also a
clear separation between early and late types, in all environments, in the
stellar mass vs. V-z observed color plane, with bulge-dominated galaxies sitting mainly
in the red sequence and disk-dominated ones populating the blue cloud.  Furthermore,
they fit and discuss extensively the different slopes of the blue
cloud (late-type galaxies) and the red sequence (early-type galaxies)
distributions. We see similar characteristics in our data but, in our
opinion, this is an effect mostly due to the sample mass
completeness. As it is clear by looking at the mid column of
Figure~10, the pronounced slope of the late-type galaxies distribution,
compared to the bulge-dominated galaxy population, is mostly drawn by the mass
completeness: red objects with such low mass, if existing, would have
an I band magnitude fainter than 24.  

Nonetheless, their main results
are in excellent agreement with the one presented in this work at
z$\approx$0.7, in spite of the different ways used to estimate galaxy
environmental properties.

\begin{figure*}
\begin{center}
\includegraphics[angle=0,width=0.85\linewidth,bb = 19 145 580 710]{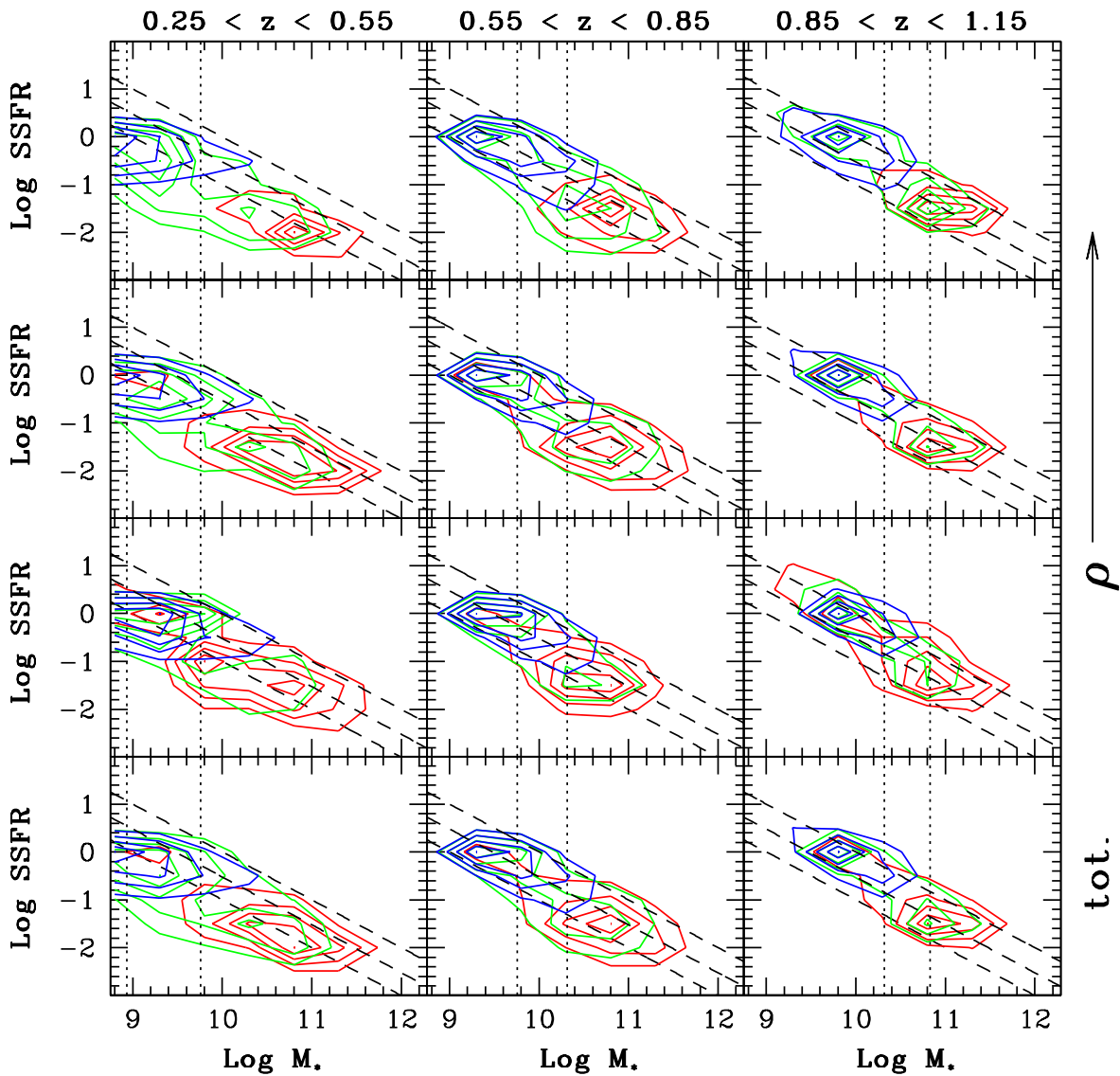}
\caption{The SSFR as a function of stellar mass and morphology for three different redshift bins and in different environments. Contours define the morphological classes and are drawn as explained in the text. Tilted dotted dashed represent constant star formation rates of 1, 3 and 10 M$_\odot$yr$^{-1}$ and they are supposed to guide the eye. The bottom raw shows the total value. The two vertical dotted lines show the mass completeness values at the upper and lower edges of each redshift bin.}
\label{ssfrl}
\end{center}
\end{figure*}

\subsection{The specific star formation rate}
\label{sec_ssfr}

One way to explore the contribution of star formation to the stellar
mass growth in galaxies of different mass, is to study the redshift
evolution of the specific SFR (SSFR), which is defined as the SFR per
unit stellar mass.  A number of works in the last
years~\citep[e.g.][]{brinchmanneellis2000,feulner:3,feulner2005,bauer2005,iglesias2007,zheng2007,cowie2008,damen2009, dunne09,santini09, pannella09} have focused on the SSFR redshift evolution, consistently finding an increase of the mean SSFRs with redshift. 

Following \citet{mad_poz_dick1}, the SFR of individual objects can be
estimated from the rest-frame UV luminosity as ${\rm
SFR}_{2800}=1.27\times10^{-28}\times L_{2800}$ in units of
M$_\odot$yr$^{-1}$, where the constant factor is computed for a
Salpeter IMF. The SFR of each galaxy is corrected by the dust
attenuation obtained for the SED fitting using the extinction curve
of~\citet{calzetti1994}.

In Figure~\ref{ssfrl} we present the SSFR as a function of stellar
mass and morphology, in three redshift ranges up to redshift 1.15, and
in different environments. The Figure shows contour isodensity levels
drawn as in Figure 10. 

Before discussing in some detail Figure 11, we would like to first discuss some evident similarities with 
Figure 10. The rest-frame U-B color is bracketing the Balmer break of a galaxy spectrum, hence is mostly sensitive to the age of the galaxy stellar population~\citep[e.g.][]{gallazzi05}. To convert an age in a star formation rate, one needs, to first order and neglecting the dust attenuation correction, to normalize it to the underlying continuum emission or to the galaxy stellar mass {\it i.e.}: $Log SFR \propto - (U-B) + Log M_*$. 
Taking this into account one can, again roughly speaking, link the two plots with the following conversion:  $Log SSFR \propto - (U-B)$.  With this simple conversion in mind, it is easier to understand why there are so many similarities between the two Figures.

In both Figures the red and blue clouds are very well
separated, at all redshifts and in all environments. This
is more pronounced in high density environments, while going to lower
density regions there is clear trend of the low mass early--types to
head toward high SSFRs, or bluer U-B color, values. These galaxies are young blue bulge-dominated 
galaxies with high star formation activity and stellar masses of order a few times
10$^9$ solar masses.

The intermediate--type galaxies fall in between, again confirming
their nature of transition objects. It is worth noting though that
while they overlap with the low mass star forming late-type population at
all redshifts and in all environments, they tend to leave the high
mass end at progressively lower redshifts.

Coming more specifically to Figure 11, the location in 
the SSFR vs stellar mass plot of the maximum galaxy density of the early-type population,
is increasing its star formation by not more than a factor three over
the explored redshift range, while the blue cloud is moving by approximately 
a factor ten in the same redshift range.  The two clouds are
pretty much segregated in mass: while late-types are always dominating at
low masses, the early and intermediate type objects completely
dominate the high mass tail. This mass segregation gets definitely
stronger in denser environments.

The upper envelope of the SSFR is running essentially parallel to
lines of constant SFR, and it is shifting to higher SFRs with
increasing redshift, as it was already noted in earlier work. Note
that this upper envelope is partly affected by a selection effect:
heavily dust obscured starbursts cannot be detected in our sample.
  
Furthermore, it is evident that the most massive galaxies have the
lowest amount of star formation per unit stellar mass up to the
highest redshift probed, and hence they cannot have formed the bulk of
their stars in this redshift range. This pushes the bulk of their star
formation at earlier cosmic epochs, in agreement with the
downsizing scenario~\citep{cowie1996}.

\citet{caputi08} analysed a sample of 609 sources in the COSMOS field,
with both optical spectra from the zCOSMOS-DR2 and a MIPS
24$\mu$ photometric measurement available.  In their study they
compared different kinds of star formation rate indicators, {\it i.e.}
UV light, line emissions and IR luminosity, thus allowing an
investigation of the dust extinction affecting these diagnostics. As a
main result, at least for what concerns our work here, they find that
the Calzetti et al. (1994) attenuation law works well for de-reddening
both the UV light and the H$_\delta$, H$_\alpha$ line fluxes. Although
the median $A_V$ of our sample, as derived from the SED fitting, turns
out to be somewhat lower than the one they find for the MIPS 24$\mu$
selected sample, we note that this might be partially due to a
selection effect of the \citet{caputi08} sample.  We postpone a more
careful analysis of this issue to a future work, which will make use
of the recently available Spitzer Space Telescope mid--infrared
observations~\citep{sanders2007}.

In Figure \ref{ssfrl} it is also worth noting the evolution of the
SSFR as a function of the environment. At a first glance, there is not
much difference between the different environments. Nonetheless, at
all redshifts there is a tendency of the early--type sample to present
a plume toward the high SSFR zone, which is very pronounced at lower
environmental densities.

\section{Discussion and conclusions}
\label{sec_conclusions}

In this work we have studied the evolution of the stellar mass content
of disk-dominated and bulge-dominated galaxies with respect to the total mass budget up to
$z\sim1.2$, and its dependency from the local environment.

As expected, we find that early--type objects always dominate the
high--mass end, while at progressively lower stellar masses late--type
objects increase their contribution to the total mass density.

At each redshift, a {\it transition mass} can be identified as the
stellar mass where the transition from a bulge--dominated to a
disk--dominated stellar mass budget takes place.  Based on the good
statistics available in this work, we are able to confirm that the
morphological mix at the high--mass end evolves with cosmic time, with
the transition mass increasing with redshift. At $z\sim 0.7$ we find a
transition mass approximately consistent with the local value
$M\approx5\times10^{10}$M$_\odot$, while at $z\sim1$ the late-types and early-types
contributions become comparable at $M\approx1\times10^{11}$M$_\odot$.

The stellar mass function of disk-dominated galaxies is consistent
with being constant over the redshift range [0.3-1.2], with number
densities declining by more than an order of magnitude in the mass
range Log M$_*$ = [10.6 -- 12]. On the other hand, the stellar mass
function of bulge-dominated systems shows a pure normalization decline
with redshift. Such different behavior of the stellar mass functions
of late-type and early-type galaxies naturally sets the redshift evolution of the
transition mass.

The slope of total number density evolution with redshift is consistent, 
within the errors, with being the same in all mass ranges,
at least up to the common redshift completeness. This means that the
galaxy stellar mass function, in the redshift range and stellar mass range explored, does not 
evolve in shape but only in normalization with redshift~(see also \citet{marchesini2008}
for similar conclusions at higher redshifts). According to our data we do not see any
differential evolution between the high mass and the low mass end of the galaxy stellar mass function.

We have identified in the photometric redshift distribution a large
scale structure underdensity at z$\approx$ 0.5. It gets more and more
significant with increasing stellar mass.  
The morphological evolution in this underdense redshift bin is {\it
delayed} at all stellar masses: the early-type fraction, both with respect to the total
number density and to the late-type number density, is much lower than
those in the contiguous redshift bins, in agreement with the
expectations of the morphology-density relation.  We notice that the
morphological mix in this underdense redshift bin is, at all masses
probed, very close to the morphological mix at z$\approx$1.1, which
has a comparable total comoving number density of objects.

We further explored the redshift evolution of the galaxy stellar mass
function split by morphology and by environmental density.  In
overdense and medium-dense environments, the mass function shape is in
good agreement with local determinations.  However, we find that the
slope of the mass function is anti-correlated with the environmental
density: the less dense is the environment the more steep is the mass
function. This trend has also been found locally by \citet{baldry2006}.

In overdense regions, at all redshifts probed, a very massive early-type
population seems to exist, which overcomes the local {\it field} mass
function determination: we might be witnessing the
appearance of giant elliptical galaxies in the highest density
regions.

The morphologically-split mass content evolves with both redshift and
local density, with a striking feature: at different redshifts, the
morphological mix and the transition mass appear to mainly depend on
the local number density.

We have presented both the fraction of early-type
galaxies with respect to the total galaxy population and the ratio between
early to late--type objects at different redshifts and
for different environments, finding that a morphology--density
relation is already well in place at redshift 1.

 We have
found that early and late--type galaxies in the stellar mass vs. SSFR
plane (or stellar mass vs. rest-frame U-B color) are well separated
at all redshifts and in all environments. The intermediate type
galaxies fall in between, confirming their nature of transition
objects. At increasing redshift, the peak of the early--type galaxy
distribution in such a plane is moving toward higher SSFR by a factor
$\leq 3$. At the same time, the peak of the late--type galaxy
distribution shifts by approximately a factor ten in the same redshift
range. 

The early-- and late--type galaxy populations exhibit a
significant segregation in mass: late-type galaxies dominate at low masses
while early--type and intermediate objects dominate the
high mass tail. Therefore, early--type and intermediate galaxies drive
the evolution of star formation present in massive ($\ge 7 \times
10^{10} M_\odot$) galaxies, while lower mass disk-dominated galaxies are mostly
responsible for the global decline of the cosmic star formation from
redshift one to the local Universe.  

While in general this picture
seems to be quite similar in all environments, in low density regions
there is a population of relatively massive, early-type galaxies,
having high SSFR and blue colors.

We also explored, with the highly homogeneous dataset available, the
age of the massive early--type galaxy stellar populations as a
function of environment. The massive early--type galaxies have similar
characteristic ages, colors, and SSFRs, hence a similar formation
redshift independently of the environment they belong to.  The age
distributions in different environments show a difference in the mean
ages (neglecting the age-metallicity degeneracy) which is in a good
agreement with published results. 

Finally, in agreement with previuos work, we have found that the galaxies hosting the oldest
stars in the Universe preferentially belong to the highest density regions.

\acknowledgements We thank the anonymous referee for constructive comments which helped us to improve significantly the presentation of our results. MP and VS are grateful for support from the Max-Planck Society and the Alexander von Humboldt Foundation. Part of this work was supported by the German \emph{Deut\-sche For\-schungs\-ge\-mein\-schaft, DFG\/} SFB 375. MP thanks Gabriella De Lucia, Daniele Pierini and Stefano Zibetti for a careful reading of the manuscript and for many insightful comments and suggestions. We gratefully thank the entire COSMOS collaboration for the huge amount of work and high quality data released to the whole community. The HST COSMOS Treasury program was supported through NASA grant HST-GO-09822. More information on the COSMOS survey is available at http://www.astro.caltech.edu/cosmos. The National Radio Astronomy Observatory is a facility of the National Science Foundation operated under cooperative agreement by Associated Universities, Inc. This research has made use of the NASA/ IPAC Infrared Science Archive, which is operated by the Jet Propulsion Laboratory, California Institute of Technology, under contract with the National Aeronautics and Space Administration.

\appendix 

\section{The morphological analysis}

\label{app:simul}
In the following we describe the analysis aimed at quantitatively
determine the morphological properties of the galaxy sample we use in
this work. The morphological analysis is based on the ACS imaging of
the COSMOS field in the F814W passband, and was performed on all
galaxies in our sample down to a limit of F814W=24.

We explain below the details of the adopted procedure and the tests
that were performed to assess the robustness of our results.

\subsection{Fitting  S\'ersic  profiles with GIM2D} \label{sec_gimtwod_general}

We use the package GIM2D ~\citep{simard1999} to fit PSF convolved
\citet{sersic1968} profiles to the two-dimensional surface brightness
distribution of each galaxy:

\begin {equation}
\Sigma(r)=\Sigma_e \mbox{ e}^{-\kappa \left[({\frac{r}{r_e}})^{1/n} - 1\right]}
\end {equation}

\noindent where $r_e$ is the effective radius of the galaxy,
$\Sigma_e$ is the surface brightness at $r_e$, $n$ is the power-law
index~(S\'ersic~index), and $\kappa$ is coupled to $n$ such that half of the total flux
is always within $r_e$.  For $n\gtrsim 2$, $\kappa\approx 2n - 0.331$;
at low $n$, $\kappa(n)$ flattens out towards 0 and is obtained by
interpolation.  The adjustable form of the S\'ersic law has the
advantage of parameterizing, through the variable exponent $n$,
surface brightness distributions including the exponential radial
fall-off of the light profile in bulgeless disks ($n=1$), and the
classical de Vaucouleur profile typical of elliptical galaxies
($n=4$).

During the fitting process, the model is convolved with the user
specified PSF, and then compared to the input image. When fitting a single S\'ersic model, GIM2D seeks the best fitting
values for the following eight parameters: total flux, half-light
radius, ellipticity, position angle, center of the galaxy, background
level and value of the S\'ersic index.

\subsection{PSF influence on the morphological parameters} \label{secsec}

The accurate determination of the image PSF is important for the
proper determination of the morphological parameters.  The PSFs used
to convolve the profiles were obtained for each individual ACS tile by
stacking about 10 high S/N isolated stars.

Although we did not expect the outcome of single S\'ersic fits to
depend strongly on small variations of the PSF, we carried out some
tests with varying PSFs on a few COSMOS tiles to assess the influence
of the PSF on the retrieved morphological parameters. Such tests were
performed by using: i) a single universal PSF obtained by stacking all
the COSMOS high S/N isolated stars; ii) the PSF we derived for the FDF in Pannella et al. (2006), 
which was observed in the same filter F814W and at
almost the same depth as COSMOS. We found that the largest differences
obtained in the output parameters were of order 3--5 \%, meaning that
single \sersic~fits are pretty robust against such PSF variations.

We also tested the accuracy of the PSF itself by fitting all the
point--like sources as selected with the SExtractor catalog
parameters. Specifically, we used the neural network star classifier
(CLASS\_STAR), the half--light radius, and the FWHM, to identify
point--like sources. When GIM2D--fitting such point--like sources, the vast 
majority -- but the saturated stars -- has a best--fit half--light radius 
smaller than 0.01 pixel, meaning that after the code convolves them with the
provided PSF they are indeed recovered as point--like sources.

\begin{figure}
\begin{center}
\includegraphics[angle=0,width=0.45\linewidth,bb = 19 145 580 700]{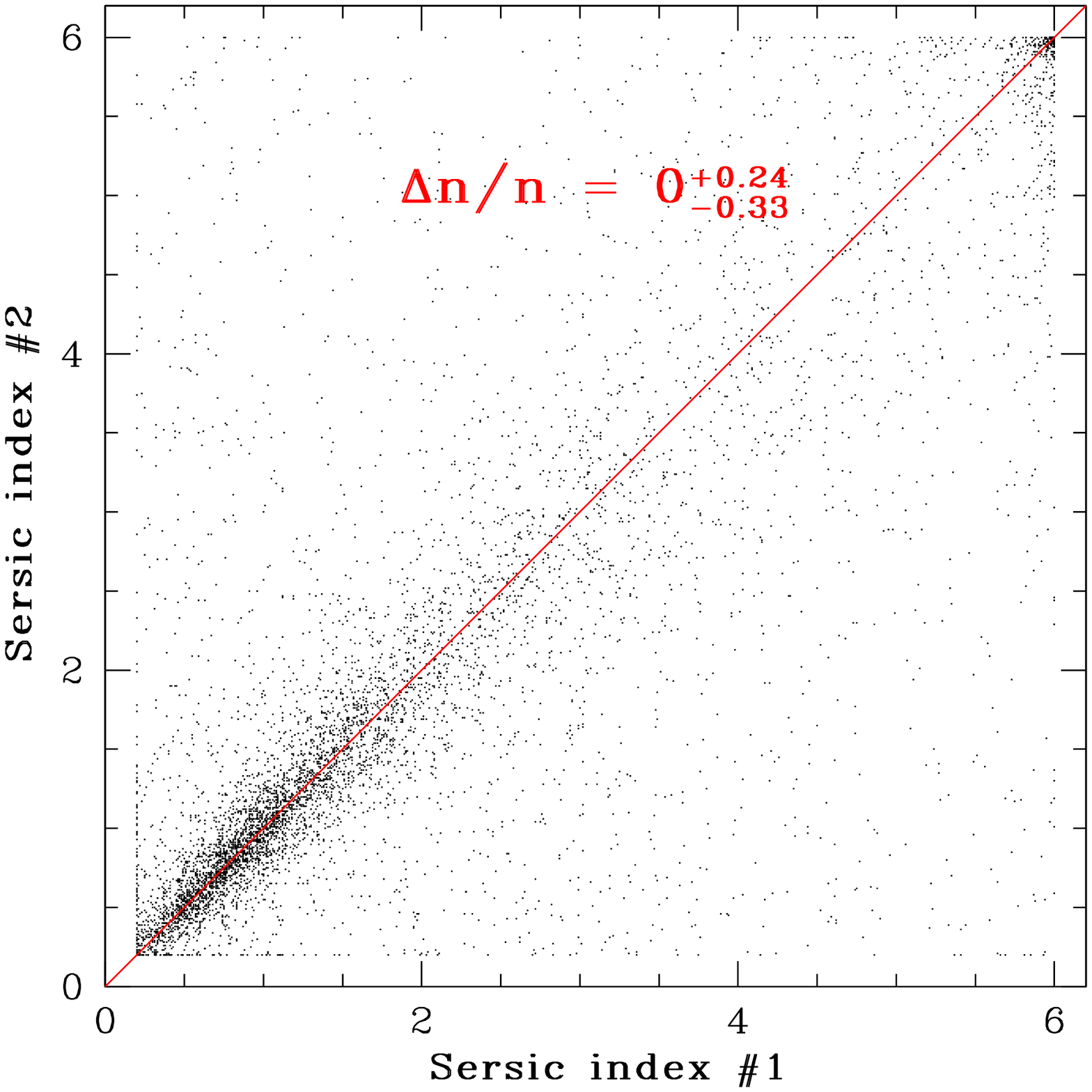}
\includegraphics[angle=0,width=0.45\linewidth,bb = 19 145 580 700]{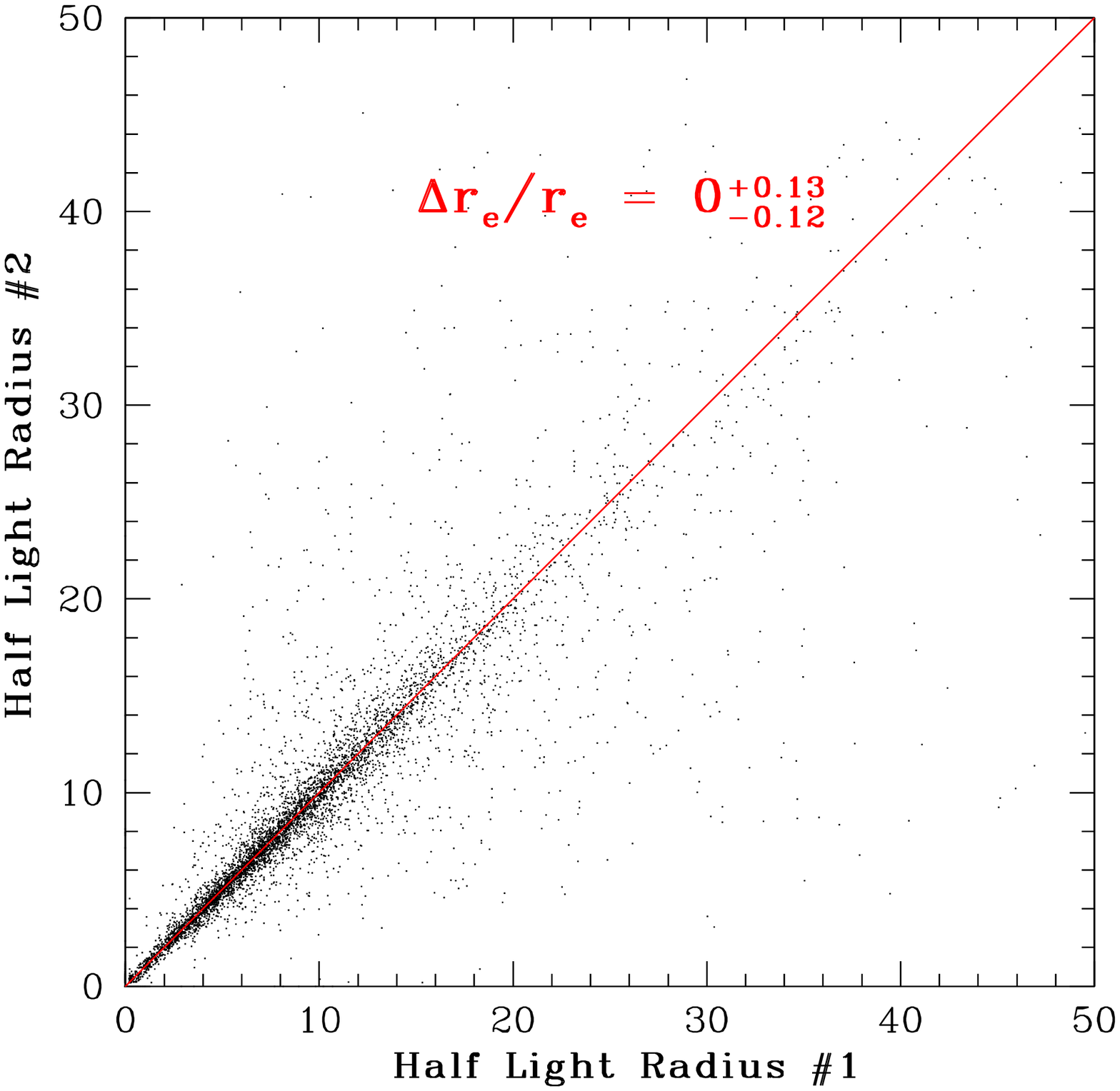}
\caption{Comparison of morphological parameters obtained by fitting the surface brightness of all sources falling in two distinct ACS tiles. For each object, the fit was performed on both images, and the resulting   \sersic~indexes and effective radii are compared in the left and right panels, respectively. The recovered median and 16th, 84th percentiles of the  $\Delta n/n$ and  $\Delta r/r$ distributions are inserted in the two panels.
 \label{nnrr}}
\end{center}
\end{figure}

\subsection{Results on the overlapping COSMOS ACS tiles} \label{secsec}

Since the COSMOS ACS tiles are slightly overlapping, we took advantage
of this feature to further test the reliability of our fitting
procedure as well as to assess many other possible systematics, like
for instance PSF variation in time and over the ACS field of view,
or photometric and astrometric distorsion.

All sources appearing in two tiles were fitted separately in each
tile, and the fit results were compared. The outcome of this test is
shown in Figure~\ref{nnrr}, where we plot one against the other the
GIM2D output (\sersic~index and half--light radius) coming from the
two fits for all such objects ($\approx$~8~000).  An overall accuracy
of $\Delta n/n \sim 0.3$ and $\Delta r/r \sim 0.1$ is obtained. Taking
into account that in the tile borders only three out of four dithers
are overlapping, and hence that the image rms is greater than in the
central part of the image, one should consider these uncertainties as
safe upper limits.

\begin{figure}
\begin{center}
\includegraphics[angle=0,width=0.45\linewidth,bb = 19 145 580 700]{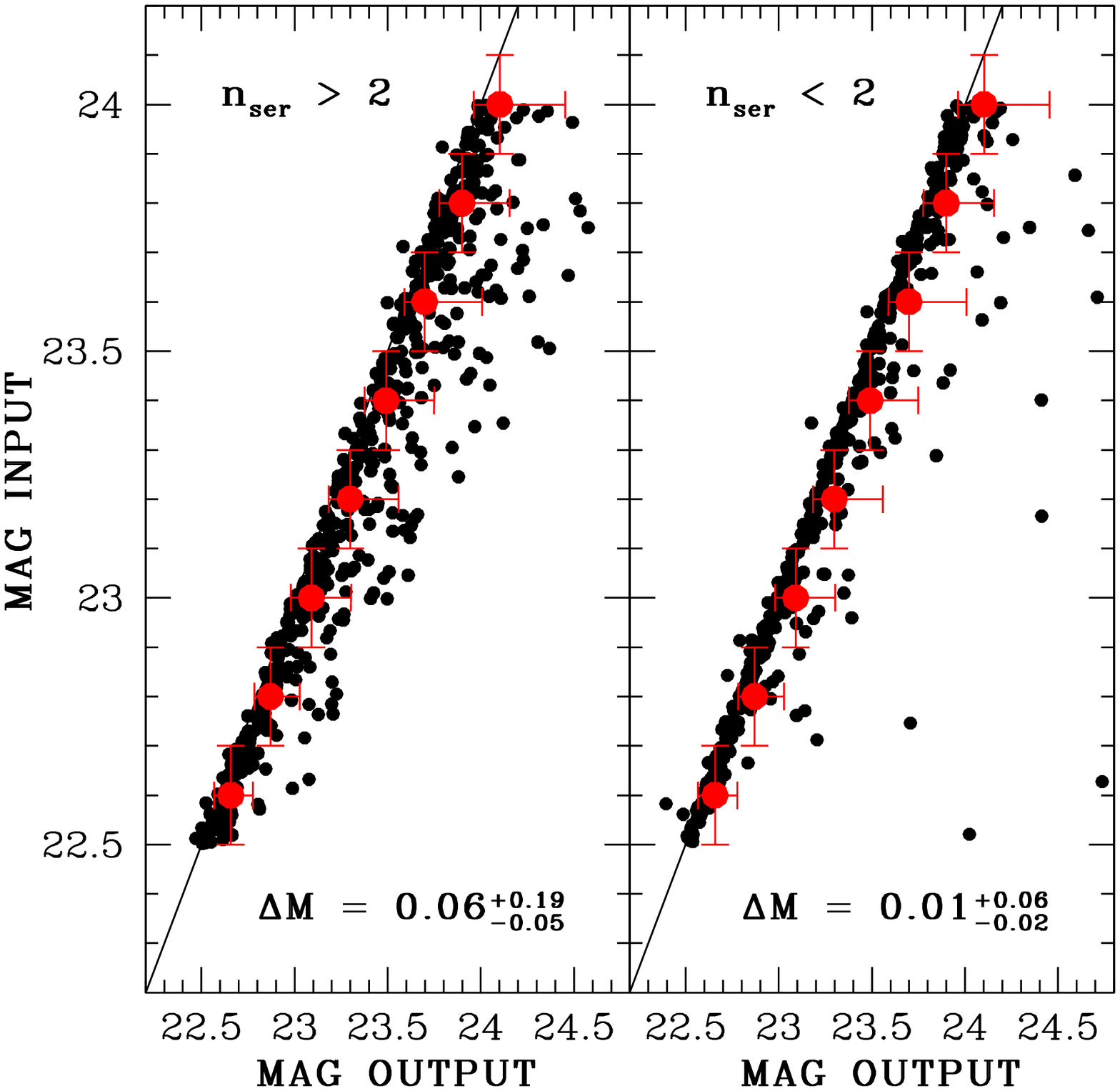}
\includegraphics[angle=0,width=0.45\linewidth,bb = 19 145 580 700]{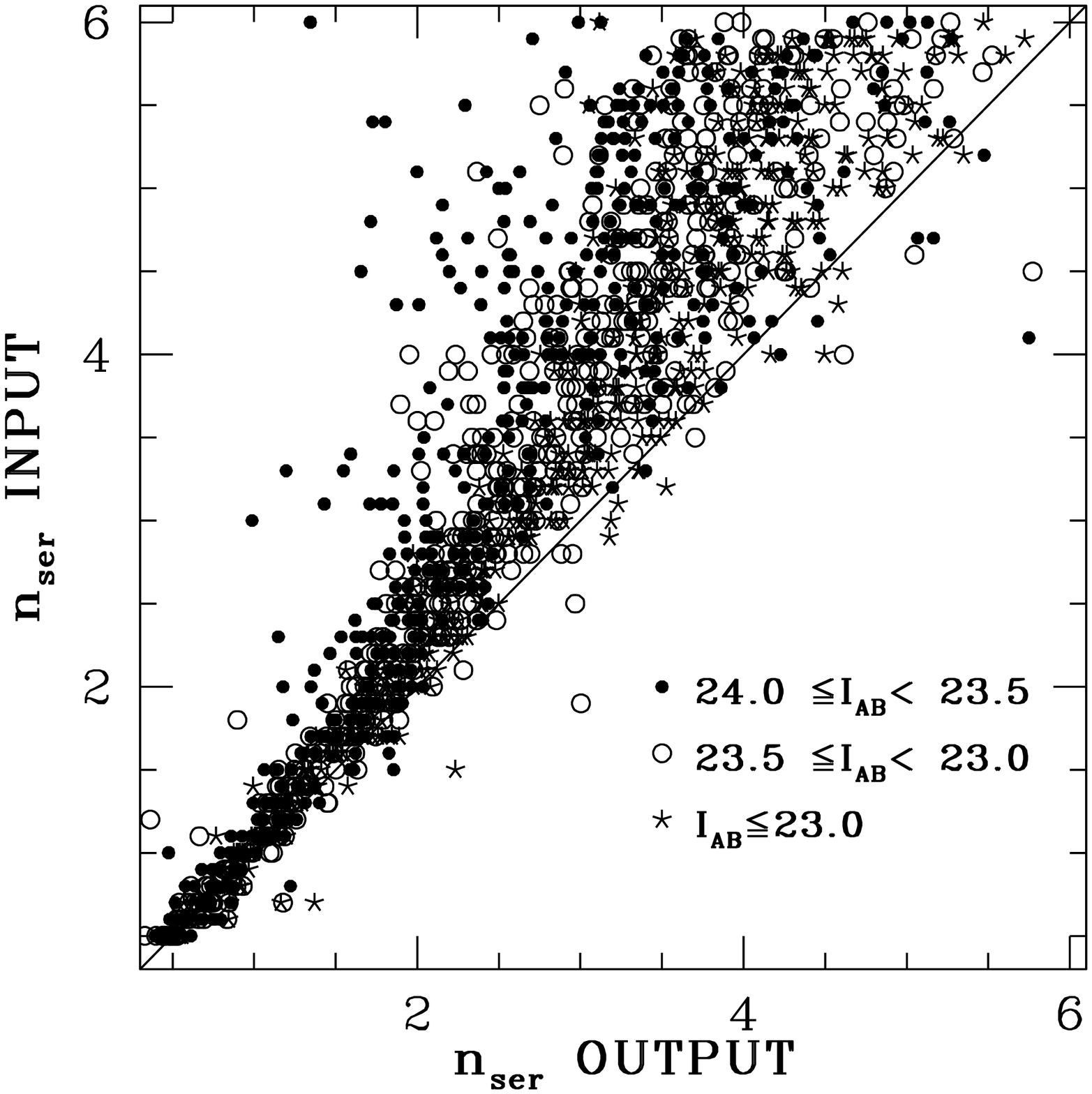}
\caption{{\bf Left}:~In the two panels output GIM2D recovered vs. input simulated magnitudes are shown for objects with  $n_{ser} > 2$ (left) and $n_{ser} < 2$ (right). In each panel the median differences with errors are shown. \label{mags}{\bf Right}:~GIM2D recovered $n_{ser}$ vs. simulated ones. High \sersic~index objects with magnitudes in the range 23/24, despite being easily detectable, fall with their extended wings under the sky surface brightness and suffer from a recovered lower value.\label{pann}}
\end{center}
\end{figure}

\begin{figure}
\begin{center}
\includegraphics[angle=0,width=0.45\linewidth,bb = 19 145 580 700]{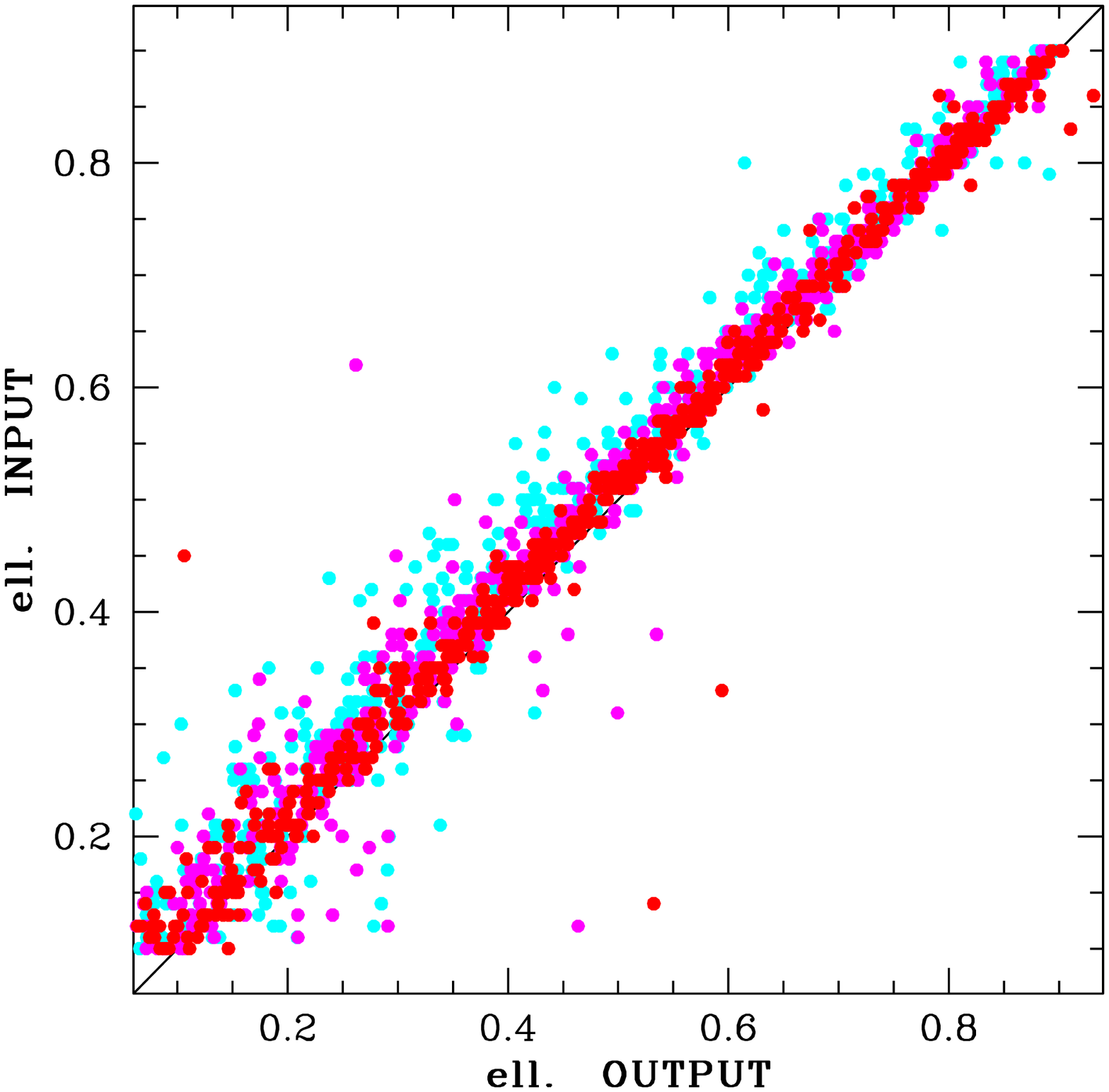}
\includegraphics[angle=0,width=0.45\linewidth,bb = 19 145 580 700]{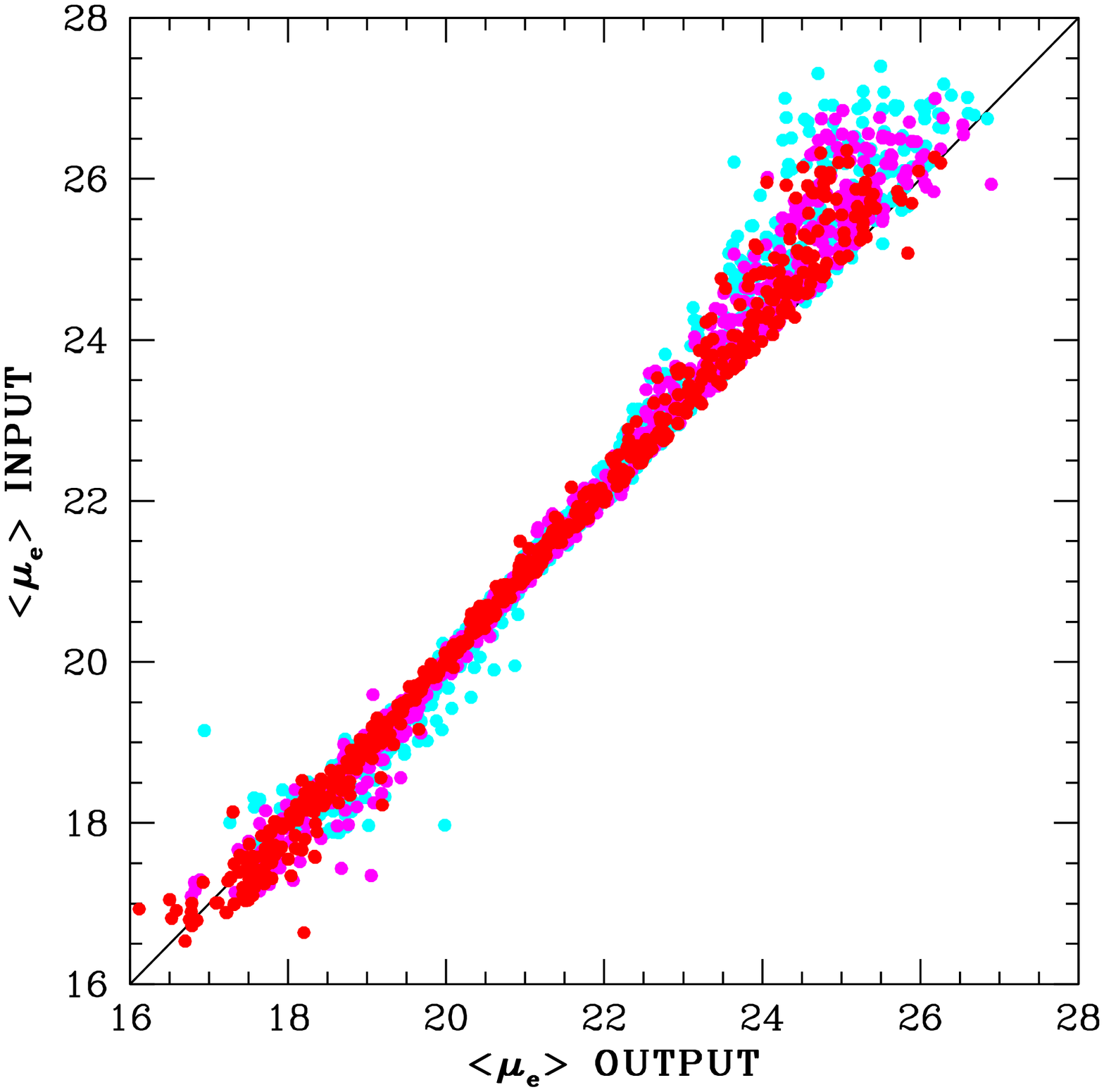}
\caption{GIM2D test results on simulated objects. The plots show the output vs. input ellipticity (left panel) and  $\mu_e$ (right panel). Points are colored according to the simulated objects total magnitudes in red (I$<$22.5), magenta (22.5$<$I$<$23.5) and cyan (23.5$<$I$<$24) \label{eeemam}}
\end{center}
\end{figure}

\subsection{Results on simulated objects} \label{secsec}

We further tested the accuracy of our analysis by running GIM2D on
simulated images. This test was performed by adding 3000 fake objects,
one at a time, to the real image, in a blank, pure--sky
region.  The simulated galaxies were generated with the GIM2D task
\textit{gsimul}, by adding on the selected pure--sky region objects
with structural parameters uniformly distributed in specified ranges
(magnitude in [22.5 -- 24.], ellipticity in [0.1 -- 0.9], position
angle in [0 -- 90], effective radius in [0 -- 60 pixel], \sersic~
index in [0.2--6]).  Each single object is first added into the image,
then it is extracted with the same SExtractor configuration file used
for the real images. It is then analyzed from GIM2D exactly as done
for the real objects. 
The input magnitudes are generally better recovered for objects with
$n_{ser} \lsim 2$ than for objects with $n_{ser} \gsim 2$, with a
median $\Delta$M = 0.01 and 0.06, respectively (see left panel of Figure~\ref{mags}).
High \sersic~index objects have very extended wings that, depending
on the total flux, can fall under the sky surface brightness, thus for
these objects a lower \sersic~index, total flux and effective radii
are usually recovered (see right panel of Figure~\ref{pann}).
We find that the profile parameters of simulated objects are well recovered down  to F814W=24 (see Figure~\ref{eeemam}). The recovered ellipticity and position angle, as compared to the input
ones, show no trends with magnitude and are generally very well
recovered (see left panel of Figure~\ref{eeemam} for the ellipticity).

\begin{figure}
\begin{center}
\includegraphics[angle=0,width=0.45\linewidth,bb = 19 145 568 700,clip]{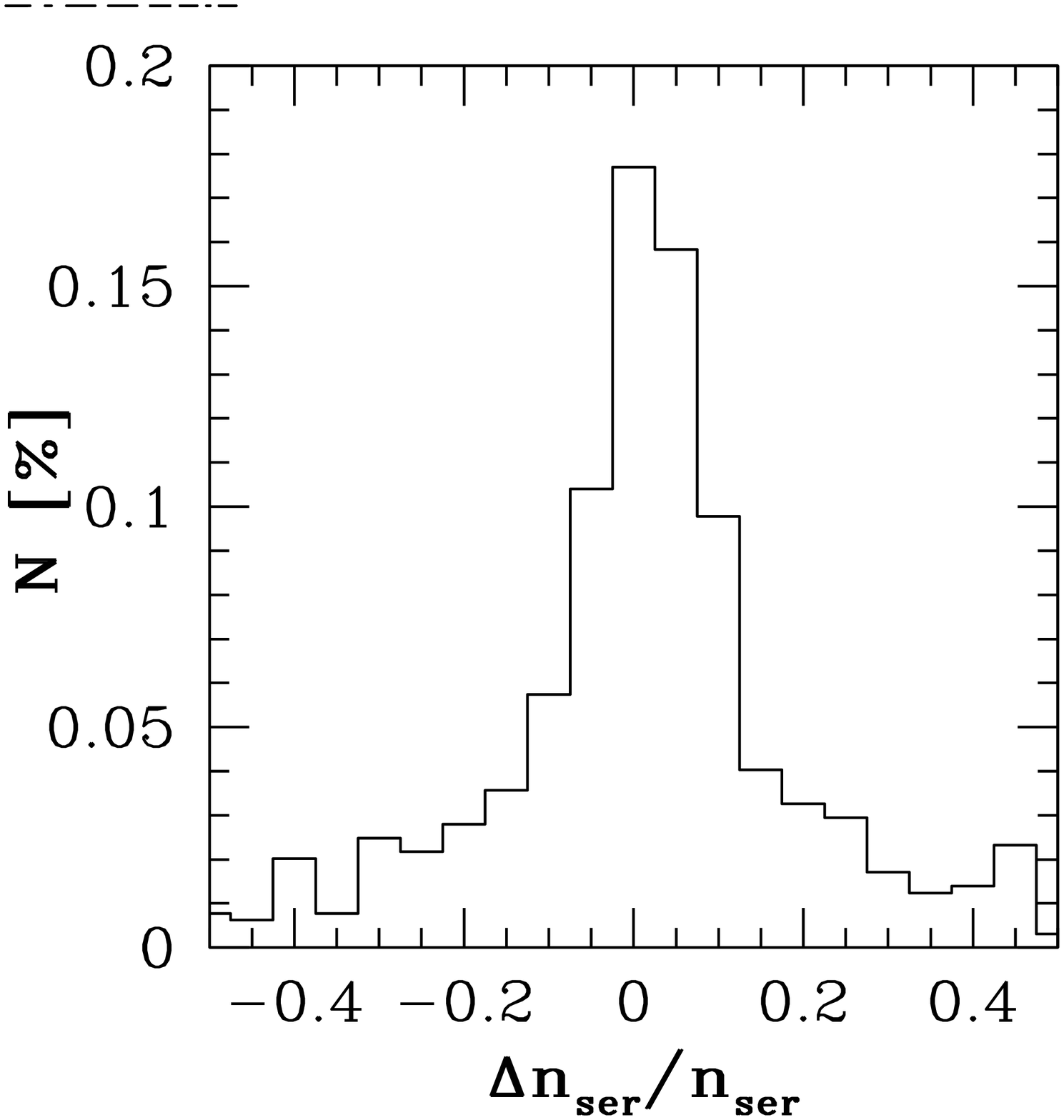}
\includegraphics[angle=0,width=0.45\linewidth,bb = 19 145 568 700,clip]{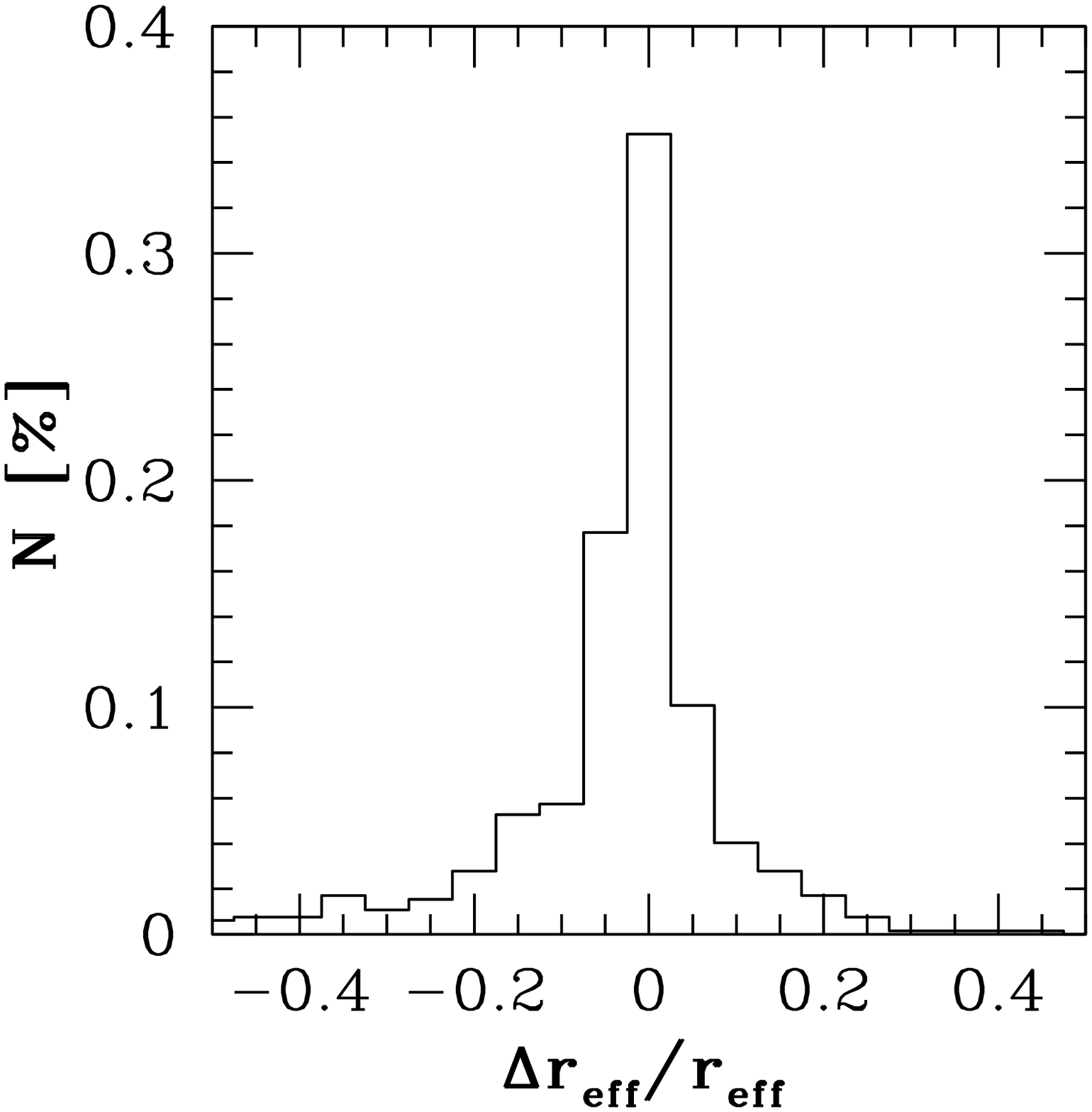}
\caption{GIM2D vs GALFIT comparison. In Figure the distribution obtained, on 4 COSMOS tile, of $\Delta n/n$ (left panel) and $\Delta r/r$ (right panel) from the two codes. The median is 0 and the width is less than 0.2 for  $\Delta n/n$ and about 0.1 for $\Delta r/r$.\label{gio}}
\end{center}
\end{figure}

\subsection{GIM2D vs. GALFIT} \label{secsec}

We tested our results by running also the GALFIT ~\citep{peng2002}
code on four of the COSMOS ACS patches. GALFIT is a 2D galaxy fitting
software package more recent than GIM2D, and it was designed to
extract many structural components from galaxy images. It is very
flexible and able to fit an arbitrary number of galaxies
simultaneously on an image, making it an ideal tool to fit neighboring
objects in a sensitive way and not being biased from external
informations such as the SExtractor segmentations used by GIM2D. 
During the fitting process, as for GIM2D, the model is convolved with
the specified PSF, and then compared to the input image.

Figs.~\ref{gio} shows a comparison between the two codes. No
systematics are present, the median differences for both the recovered
\sersic~index and half light radius are very close to zero. The
scatter is about 20\% for the \sersic~index and 10\% for the half
light radius.

One can argue that the COSMOS field is not very crowded and that
in more crowded fields one could see some systematic differences
between the two codes. Still, a similar comparison on deep ACS images
of the galaxy cluster A1689~\citep{halkola} for a sample of bright (I
$\lsim$ 22.5) red-sequence early type galaxies, beside a few obvious
cases for which GALFIT performed best, showed no significant
systematics between the two codes~(see also \citet{boris2007} for slightly
different but still similar conclusions on the comparison between the two codes). Of course these considerations apply only to single \sersic~fitting, no simulations and tests were done concerning bulge/disk decomposition.

\section{Photometric redshifts accuracy}

\subsection{Comparison with zCOSMOS DR2} 

\begin{figure}
\begin{center}
\includegraphics[angle=0,width=0.45\linewidth,bb = 19 145 580 700]{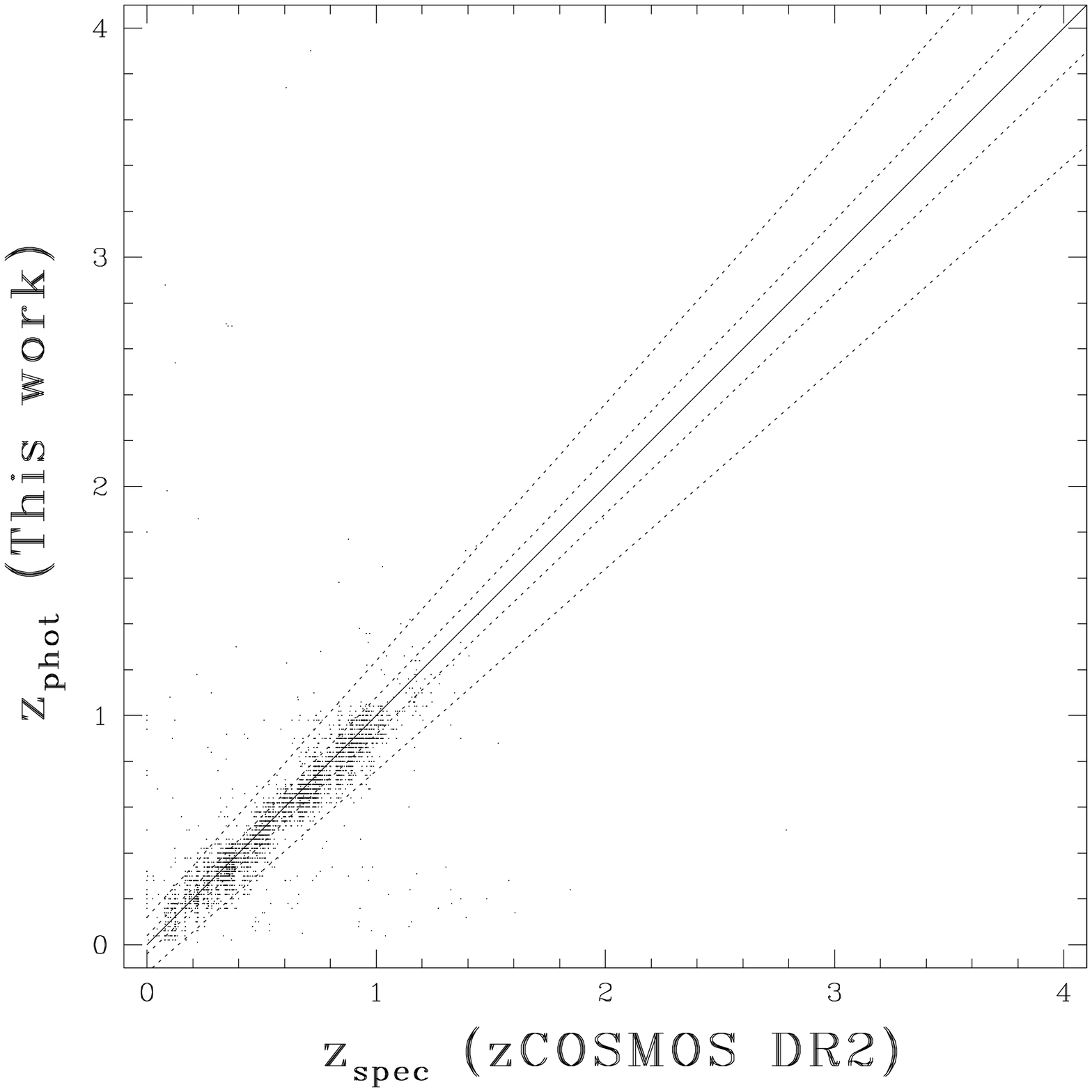}
\caption{Comparison between the zCOSMOS DR2 spectroscopic redshifts catalog and the photometric redshifts
catalog used in this work.~The dotted lines represent 0.04 and 0.12 rms values, while the solid line is the bisector, see text for details.}
\label{uno}
\end{center}
\end{figure}

In Figure~\ref{uno} we show the comparison between the photometric redshifts used in this work and the recently released zCOSMOS DR2 spectroscopic redshifts catalog~\citep{lilly2007}. 4506 objects are in common with our catalog. The distribution of $\Delta z/(1+z_{spec})$ has a median value  of 0.01 and an rms of 0.04 with only 2.5\% of the distribution having a {\it catastrophic} photometric redshift, {\it i.e.} a $\Delta z/(1+z_{spec})$ greater than 0.2. This nicely confirms the results described in Gabasch et al.(2008), which were obtained by using a much smaller spectroscopic redshifts sample.  

\subsection{Comparison with Ilbert et al., (2009)} 

\begin{figure}
\begin{center}
\includegraphics[angle=0,width=0.45\linewidth,bb = 19 145 580 700]{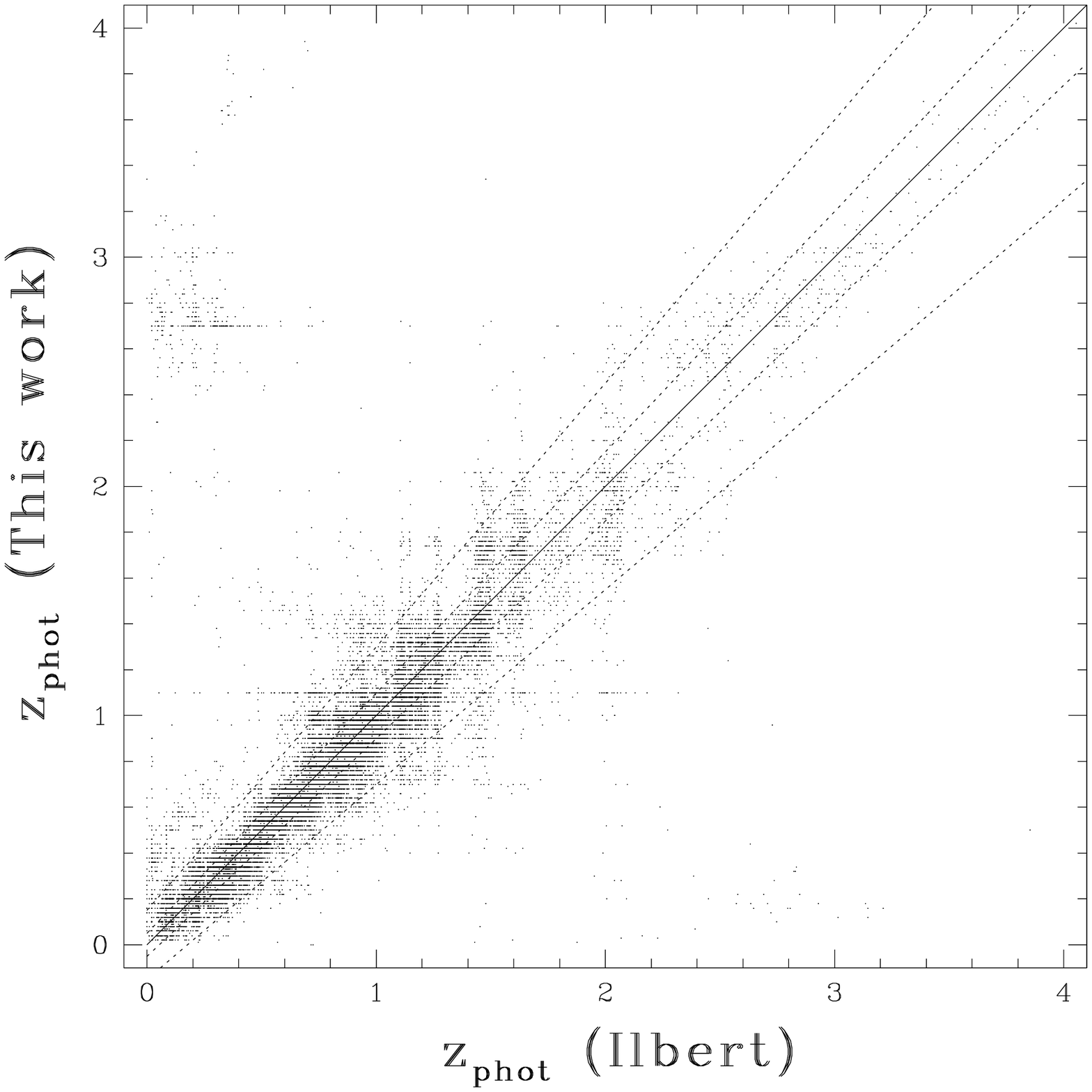}
\includegraphics[angle=0,width=0.45\linewidth,bb = 19 145 580 700]{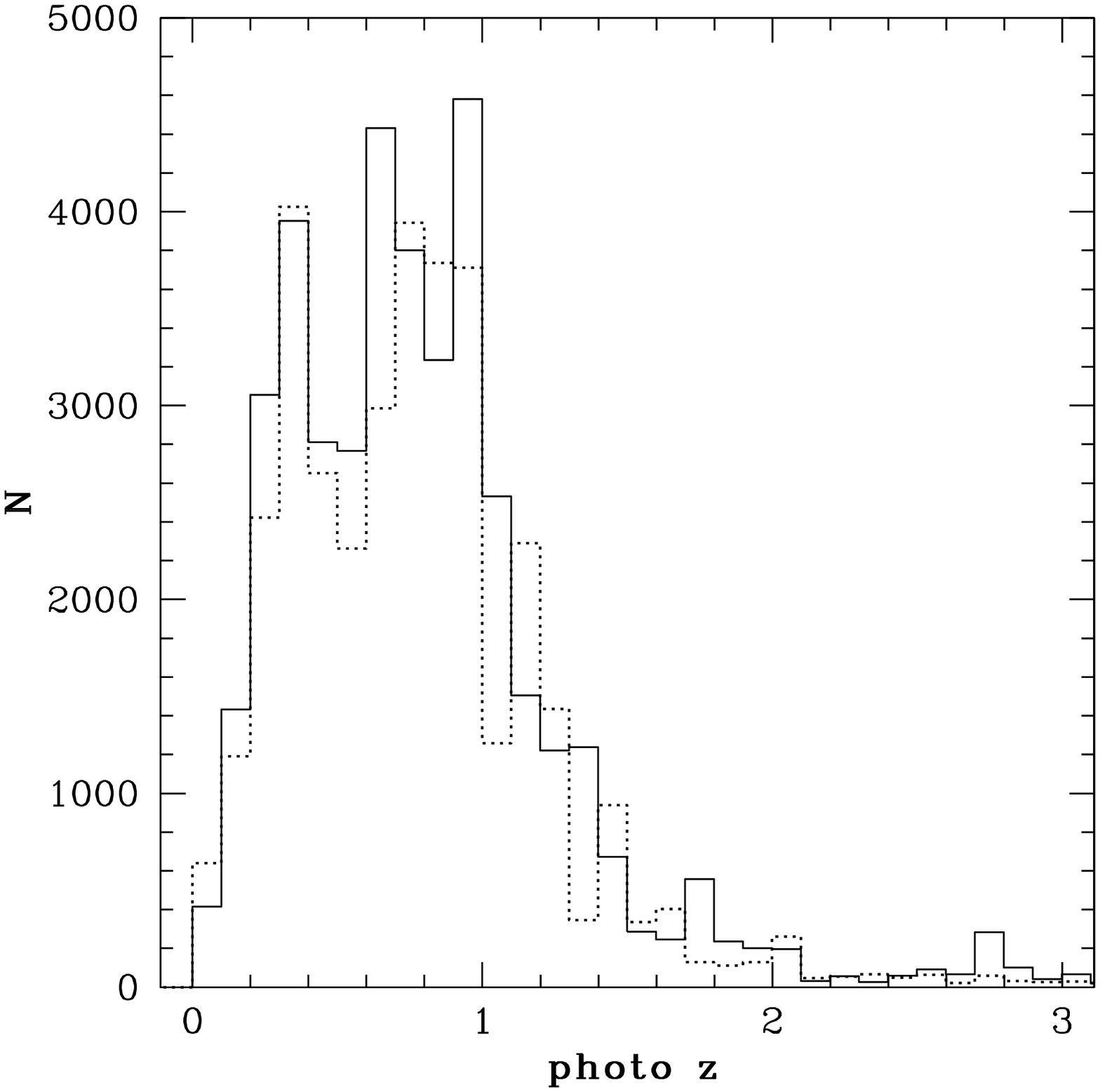}
\caption{{\bf Left}: Comparison between the photometric redshift catalog used in this work and the one recently published from Ilbert et al.~(2009). The dotted lines represent 0.05 and 0.15 rms values, while the solid line is the bisector.~{\bf Right}: Comparison between the photometric redshift distribution (solid line) used in this work~(as in Figure~1) and the one~(dotted line) from the Ilbert et al. (2009) catalog for the very same sample. }
\label{due}
\end{center}
\end{figure}

In the left panel of Figure~\ref{due} we compare the photometric redshifts used in this work with the ones from the recently released Ilbert et al.~(2009) COSMOS Legacy catalog for a sample of 40717 objects common to both catalogs. We note that not all the objects of our morphological catalog are matched because the Ilbert et al. (2009) sample does not contain the photometric redshifts for all the X-ray sources detected in the XMM COSMOS images~\citep{capp2009}. The relative accuracy between the two photometric redshift realizations is about 5\%, which is in a good agreement with the one expected from the absolute errors quoted in the two studies. Finally, only 4\% of the common objects have photometric redshifts that differ by more than 20\%.

In the right panel of Figure~\ref{due} we show the redshift distribution of our sample obtained using the Ilbert et al.~(2009) catalog~(dotted line) as compared to the one we are using in the work~(solid line, see Figure~1). We note that the two distributions agree very well, including the detection of the large scale structure underdensity  at z$\sim$0.5.
 
\end{document}